 \documentstyle[nato]{crckapb}

\def\hMpc{\mbox{h$^{-1}$ Mpc}}

\def\fp {fundamental polyhedron}
\def\uc {universal covering}
\def\rw{Robertson--Walker}
\def\ie {i.e.}
\def\frl{Friedmann--Lema\^\i tre}

\def\gtapprox{\,\lower.6ex\hbox{$\buildrel >\over \sim$} \, }
\def\ltapprox{\,\lower.6ex\hbox{$\buildrel <\over \sim$} \, }

\def\arcs{\ifmmode {'' }\else $'' $\fi}     
\def\arcm{\ifmmode {' }\else $' $\fi}     
\def\deg{\ifmmode^\circ\else$^\circ$\fi}    

\def\apj{Ap.J.}                 
\def\aanda{A\&A}            
\def\cqg{Class.Quant.Grav.}
\def\mnras{MNRAS}

\def\frtoday{le\space\number\day\space\ifcase\month\or
  janvier\or f\'evrier\or mars\or avril\or mai\or juin\or
  juillet\or ao\^ut\or septembre\or octobre\or novembre\or
d\'ecembre\fi\space \number\year}

\newcommand\joref[5]{#1, #5, {#2, }{#3, } #4}

\newcommand\epref[3]{#1, #3, #2}

\def\rinj{{r}_{\mbox{\rm \small inj}}}

\input  epsf.tex

\newcommand\zzz[2]{#1}  

\newcommand\citet[1]{\cite{#1}}
\newcommand\citep[1]{\cite{#1}}
\newcommand\citealt[1]{\cite{#1}}

\def\frprmrinj{ \begin{figure}
\centerline{\epsfxsize=8cm 
{\epsfbox[0 0 355 354]{"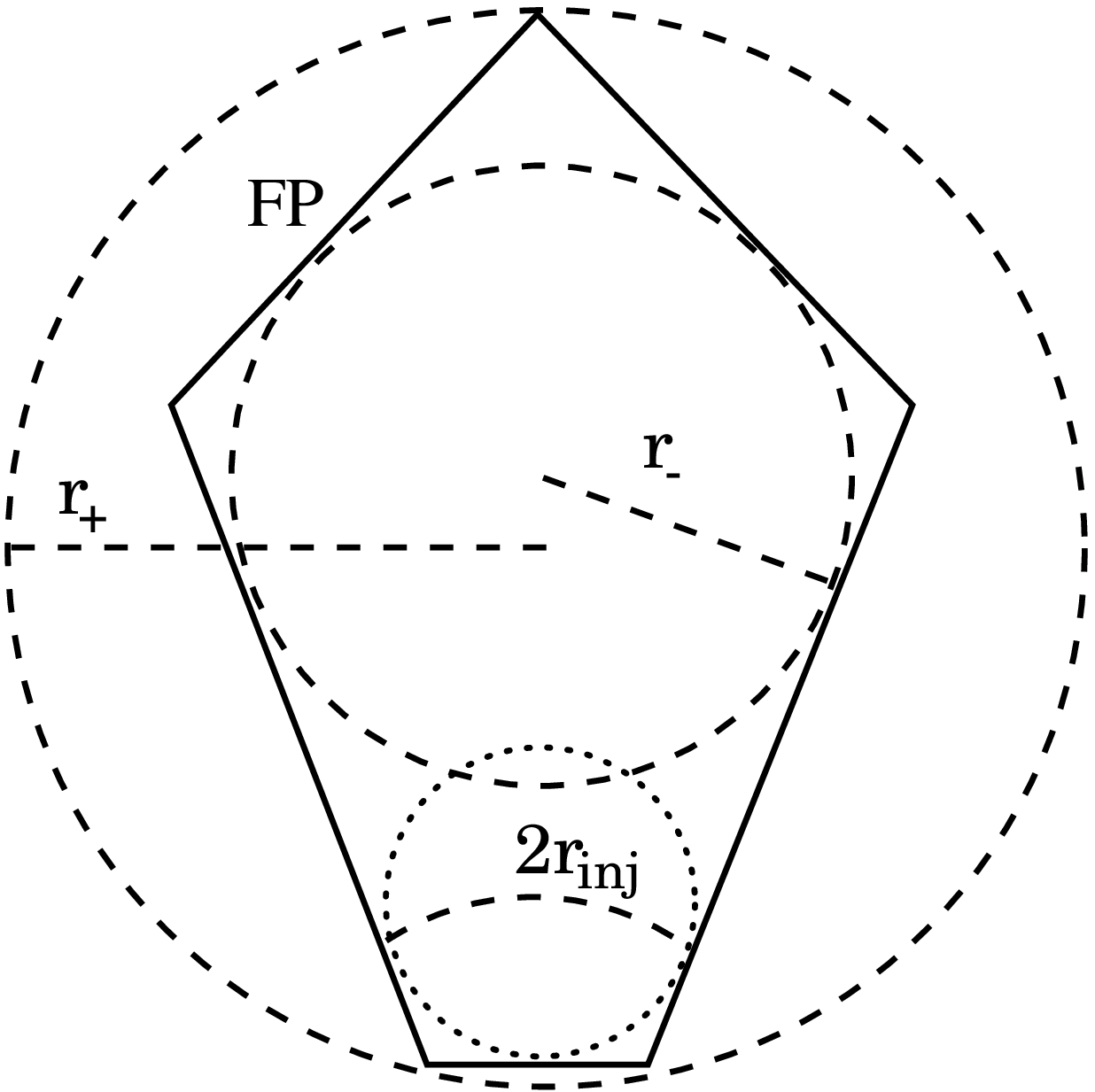"} }
}
\caption{\label{f-rprmrinj} Schematic diagram showing key
parameters representing the size of a small universe.
The terminology is that of \protect\citet{Corn98a}.
The in-radius $r_-$ and the out-radius $r_+$ are
characterised by the largest sphere inscribable in the
fundamental domain ($FD$) and the smallest sphere circumscribable
around the $FD$ respectively. The injectivity radius
$\rinj$ is half of the smallest closed (spatial) geodesic from
one topological image to another. The particular geometry shown
does not necessarily correspond to a projection of a
slice through a 3-manifold known to exist, it is for demonstration
only.}
\end{figure} }

\def\fsizes{ \begin{figure}
\centerline{\epsfxsize= 8cm
{\epsfbox[0 0 334 496]{"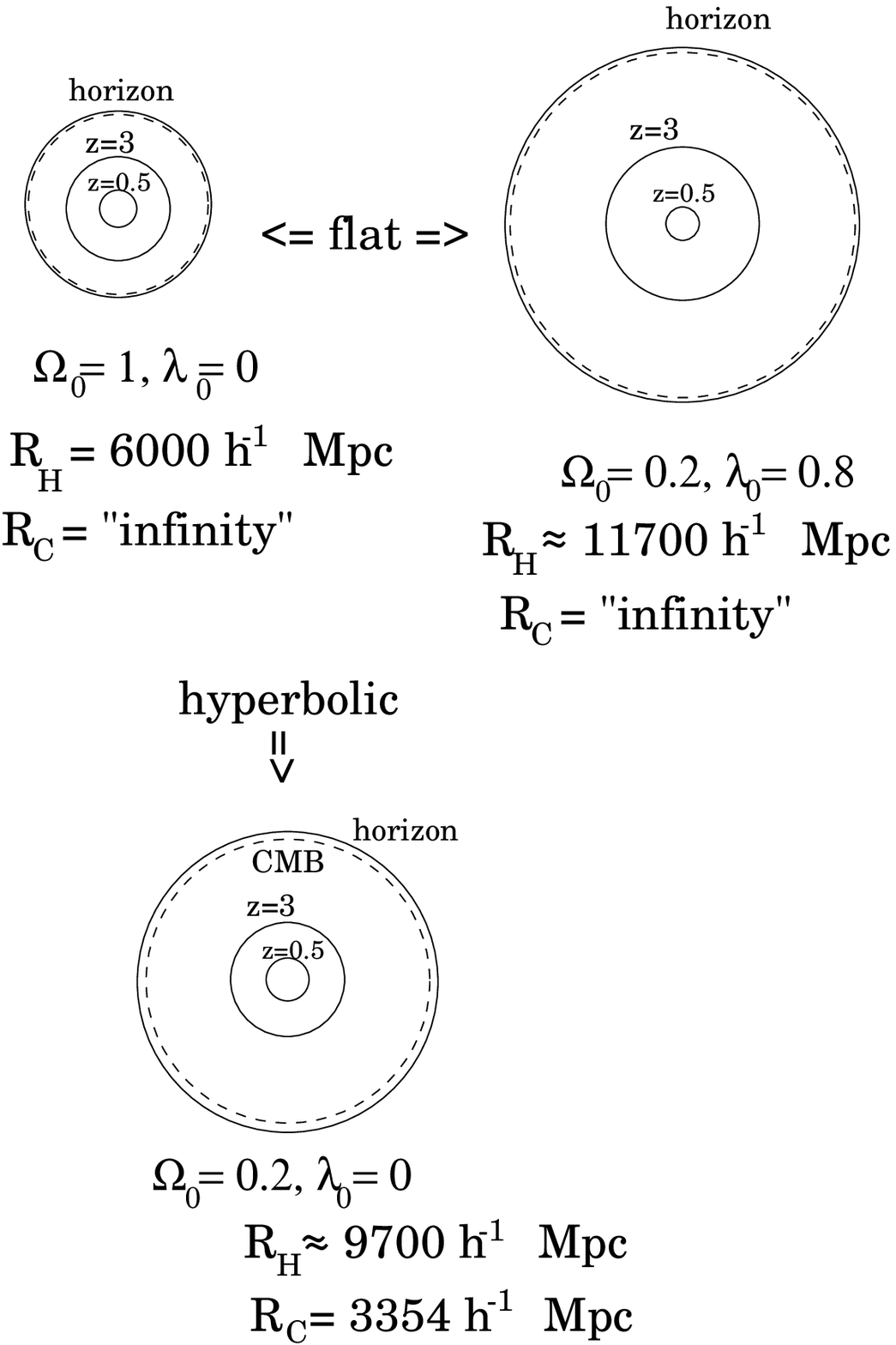"} }
}
\caption{\label{f-sizesC}
Relative sizes of the observable Universe, depending on
different options for the metric parameters
$\Omega_0, \lambda_0,$ shown in comoving
coordinates on a radially linear (proper distance) scale.
The horizon radius $R_H$ is defined by the age of the Universe and
the curvature radius is
$R_C \equiv c/H_0 (|1-\Omega_0-\lambda_0|)^{-1/2}$.
The inner circles for each choice of metric indicate scales to
which we
can see: X-ray emission from the richest clusters of galaxies
($z \sim 0.5$), large numbers of quasars ($z\sim 3$),
the cosmic microwave background ($z\sim 1100$, dashed circle).
In the flat models, the tangential distance scale is constant and the
same as the radial scales; in the hyperbolic model, the tangential
distance scale increases (more Gpc per mm on the page) as a function
of increasing radius, i.e. it is $[\sinh(r/R_C)]/(r/R_C)$ times larger
than
a constant scale, e.g. the horizon circumference is $3.1 (2\pi R_H)$.
}
\end{figure} }

\def\fcmbmethod{ \begin{figure}
\centerline{\epsfxsize=8cm 
{\epsfbox[0 0 507 557]{"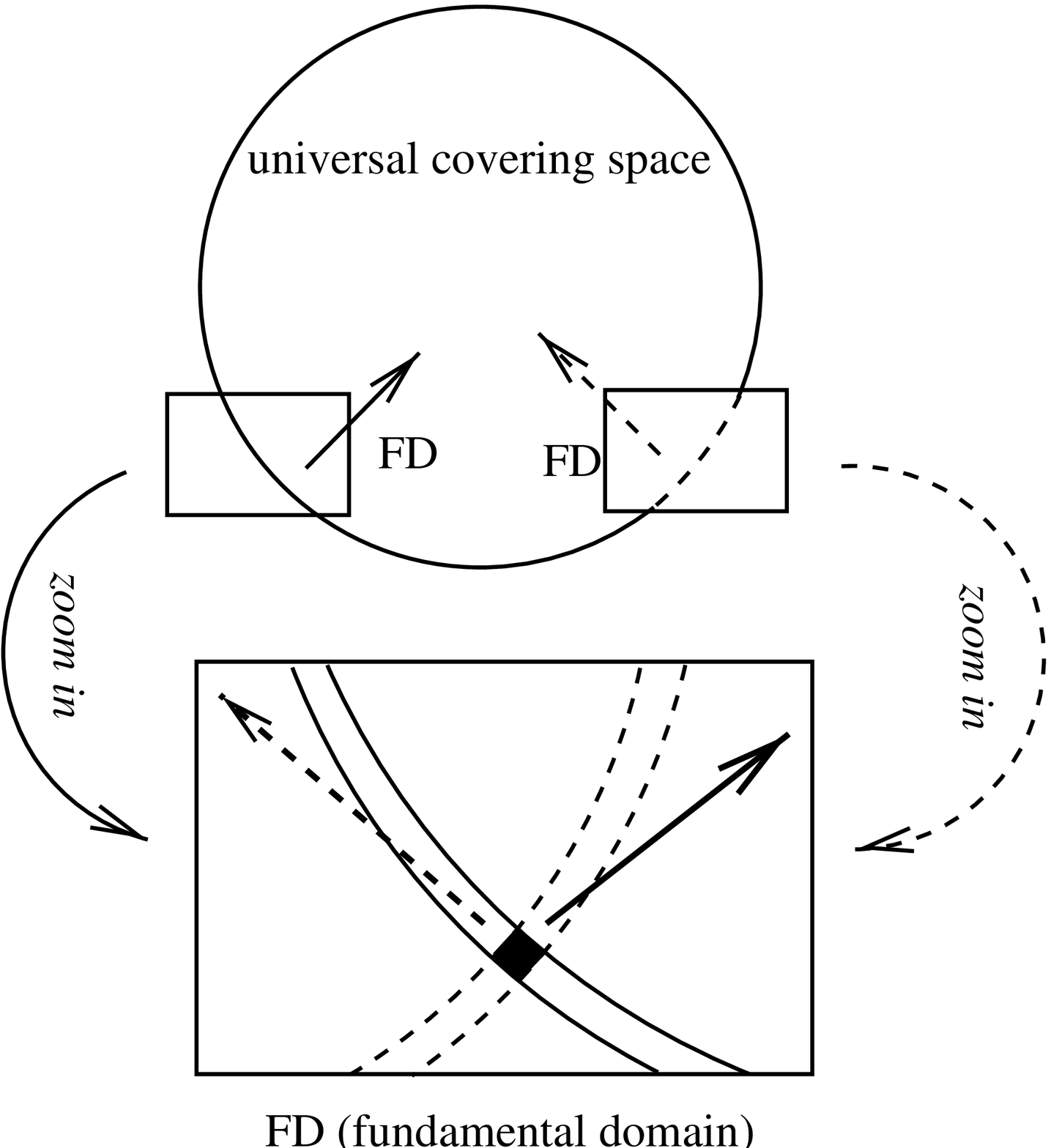"} }
}
\caption{\label{f-cmbmethod} Schematic diagram showing
basic principle of CMB methods. A single point of physical
space emits radiation isotropically at the
recombination epoch, but particularly
in two different directions such that after crossing the
Universe several times, the photons arrive at the observer.
The upper half of the figure is from the point of view of the
covering space, with several copies of the $FD$, while
the lower half is a zoomed-in more physical point of view, in a single
copy of the $FD$. The arrows indicate photons' paths.
The ``object'' which is to be seen as multiple topological images
is the intersection of the two thin sections of the shell of
last scattering, i.e. the intersection shaded in black.
}
\end{figure}
}  

\def\fweeks{
\begin{figure}
\centerline{\epsfxsize=15cm
\epsfbox[-5 539 613 726]{"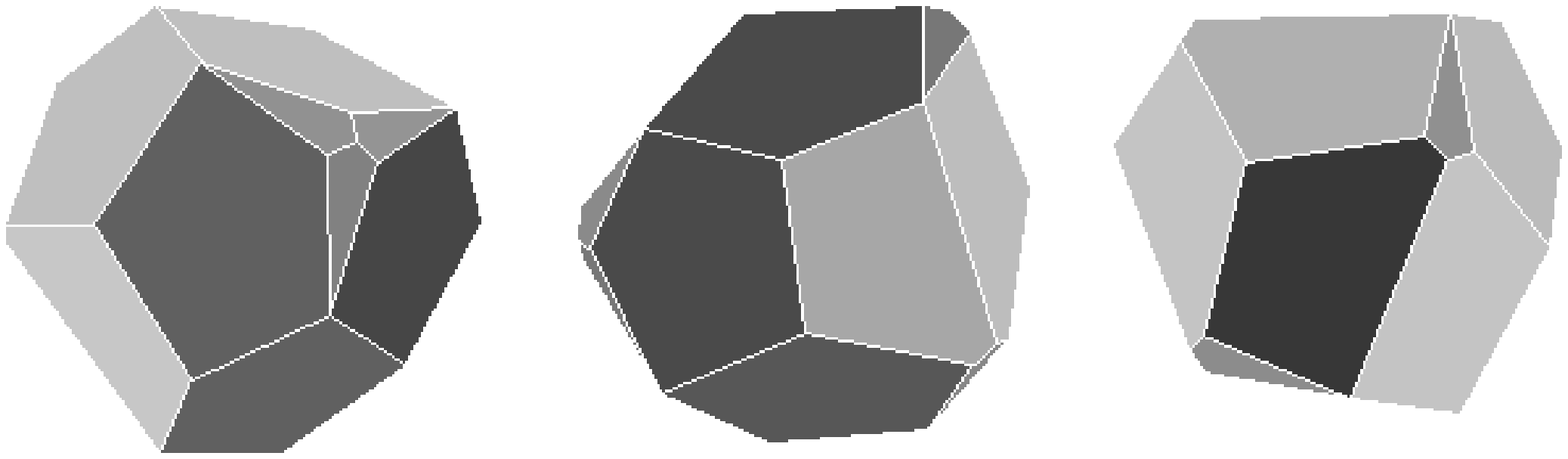"}} 
\caption{Three views of the fundamental domain of the
Weeks manifold}
\label{weeks}
\end{figure}
} 

\def\tobsvn{
\begin{table*}[htb]
\caption{\label{t-obsvn} Summary of the most recent methods
and observational results. See text (\S\ref{s-methods}) for details.
Author abbreviations are by authors' surnames' initials and year.}

\begin{tabular}{cc | c c c }
\hline
\multicolumn{2}{l}{{\bf (1) Methods:}} \\
\multicolumn{1}{l}{3D:} \\
 clus opt & cosmic crystallog. & LLL 96 \\
 clus Xray & brightest cluster & RE 97 \\
 QSO's& local isom. search & & R 96 & \\
\hline
\multicolumn{1}{l}{2D:} \\
 CMB & ID'd circles & & & CSS96/98 \\
     & $C_l$ --- cutoff & & & many \\
     & correlation fn & & & BPS98 \\
\hline
\multicolumn{3}{l}{{\em Ideal object:}} \\

& no Evoln & monotonic E & strong E & weak E? \\
& zero pec velocity& prob. small & prob. small & N \\
& isotropic emitter& Y (nearly) & N & Y/N \\
& seen to large $z$& Y ($\kappa_0 < 0$), N (o.w.) & Y & Y \\
& seen over large vol& N & Y & Y (sph shell) \\
& seen to $|b^{II}| \ll 20\deg$ & N & N & N \\

\multicolumn{3}{l}{{\em Assumptions on $\kappa_0$, $\{g_i\}$, ideal=
none:}}  \\
&&     none & $\kappa_0$ (use range) & circles: none \\
&& & &                                 $C_l$: all \\
\hline \\
\multicolumn{3}{l}{{\bf (2) Constraints:}}\\
& & CC: $2r_+ \gtapprox R_H/20$ \\
& & BC: $2r_+ \gtapprox R_H/10$ \\
& & & N/A \\
\multicolumn{5}{l}{For the following special cases, but really testing
perturbation spectrum assumptions:}\\
\multicolumn{5}{l}{$\kappa_0=0,$ if $\theta(g_i,g_j)= 90i\deg$ or
$60i\deg, i \in Z$ then}\\
&&&& ($2\rinj \gtapprox R_H/2$) \\
\multicolumn{5}{l}{$\kappa_0<0,$ if $\Gamma=m004(-5,1)$ or
$\Gamma=v3543(2,3)$ then} \\
&&&& ($2\rinj \gtapprox 2 R_H$) \\
\hline \\
\multicolumn{3}{l}{{\bf (2) Specific candidates:}}\\
&& serendipitous & 2$\sigma$ implicit & ``preferable\\
&&&& to SCDM''\\
&& $\kappa_0=0$ & $\kappa_0<0$? & $\kappa_0=-0.2$ \\
&& $\widetilde{M}/\Gamma= T^2 \times {\bf R}$ & [non-orientable] &
v3543(2,3) \\
&& $2\rinj$ ($\Omega_0$) = && $ \rinj =0.95 R_H$\\

&&$965\pm5h^{-1}$Mpc ($1$)& \\
&&$1190\pm10h^{-1}$Mpc ($0.2$) &\\
&&RE97, RB99& R96 & BPS98 \\
\hline
\end{tabular}
\end{table*}
}

\begin{opening}
\title{Topology of the Universe: Theory and Observation}

\author{Jean-Pierre Luminet$^1$}
\author{Boudewijn F. Roukema$^{2,3}$}
\institute{$^1$ DARC,
Observatoire de Paris-Meudon, 5 place Jules Janssen, \\
F-92195 Meudon Cedex, France (Jean-Pierre.Luminet@obspm.fr)}
\institute{$^2$Institut d'Astrophysique de Paris, 98bis Bd Arago,
F-75.014, Paris, France\\
$^3$Inter-University Centre for Astronomy and Astrophysics,
Post Bag 4\\ Ganeshkhind, Pune, 411 007, India (boud@iucaa.ernet.in)}

\end{opening}

\begin{document}

\begin{abstract}
``One could imagine that as a result of enormously extended
astronomical experience, the entire universe consists of countless
identical copies of our Milky Way, that the infinite space can be
partitioned into cubes each containing an exactly identical copy of
our Milky Way.  Would we really cling on to the assumption of
infinitely many identical repetitions of the same world? \ldots We would
be much happier with the view that these repetitions are illusory,
that in reality space has peculiar connection properties so that if we
leave any one cube through a side, then we immediately reenter it
through the opposite side.'' (Schwarzschild 1900, translation 1998)

 Developments in the theoretical and observational sides of cosmic
topology were slow for most of the century, but are now progressing rapidly,
at the scale of most interest which is 1-10$h^{-1}$~Gpc rather than 10kpc.

 The historical, mathematical and observational sides of this subject
are briefly reviewed in this course.
\end{abstract}

\section{Introduction}

Think of a right triangle drawn on a transparency. The transparency
is a piece of an {\sl infinite}
Euclidean plane whose flat metric is expressed by
 Pythagoras' Theorem. Bend the transparency around
into a cylinder.
The latter is thus obtained by identifying two
opposite edges of the transparency.
The new surface is
locally flat (Pythagoras' Theorem stills holds
 perfectly), but it has a quite different shape. It contains closed
 geodesics. Next we can identify
 the two remaining edges and get a flat torus (this is a thought
 experiment, because the usual torus, or ``annulus", visualized as a
surface of
 revolution in 3--dimensional
 Euclidean space, has positive Gaussian curvature at some places and
 negative at others). The flat torus has still a
 locally Euclidean metric although its global shape has drastically
 changed: it is a {\sl finite} surface without borders.

 These elementary operations well illustrate the difference between
 the curvature (given by the metric)
and the topology. Both are needed in order
to know the full geometry of a two-dimensional, or more importantly,
a three-dimensional space. This applies as well to physical space,
namely to cosmological models for
describing the Universe as a whole. The shape of space is a
fundamental issue in physics. Curiously
enough, this Cargese '98 summer school on cosmology is probably the
first one to include a couple of lectures on cosmic topology. As we shall
see below, both for historical and for practical reasons, cosmic
topology has
been widely ignored during 80 years of relativistic cosmology,
except by some pioneering authors.

After an historical section 2, sections 3-4 are mostly based on the review
paper by
\citet{LaLu95} (hereafter LaLu 95). Since the mid 90's,
worldwide interest for cosmological topology has blossomed, from a
mathematical
 point of view, mostly related to
progress in the understanding of compact hyperbolic manifolds,
from a theoretical point of view,
as
well as from
an observational point of view, related to improvements of data on
the
large scale 3--dimensional distribution of cosmic objects and on the
2--dimensional structure of the cosmic microwave background.
Since these lectures are aimed to provide a general view of the
subject, only some of the new mathematical results (obtained since the
1995 revival)
will be mentioned. An incomplete list of theoretical articles related
to the possible physical origin of cosmic topology is provided towards the end
of section 2.
For
the latest developments in all aspects of the subject,
see the proceedings of the
1997 and 1998 workshops \citep{cqg98,ctp99}. Section 5 of this
lecture, devoted to the observational aspects of cosmic topology,
provides an up-date and complement to the review of \citet{RB98}.

\section{A Brief History of Cosmological Topology}

Most of this historical introduction is taken from \citet{Lu98}, see
also \citet{Stark98}.

Is space finite or infinite, oriented or not, made of one piece or not,
has it holes or handles, what is its global shape?
A common misconception is to believe that Einstein's general relativity theory
 is all one needs to answer such paramount questions.
 However, general relativity deals only with local geometrical properties of
 the universe, such as its curvature, not with its global characteristics,
 namely its topology.

 The physical extension of space is one of the oldest cosmological questions,
 going back to twenty--five centuries of cosmological modelling (e.g.
\citealt{LuLa94}).
 In the history of cosmology, it is well
known that Newtonian physical space,
mathematically identified with the infinite Euclidean space ${\bf
R}^3$, gave
rise to paradoxes such as the darkness of the night sky (see e.g.
\citealt{Har87}) and to problems of boundary conditions.

Wondering about absolute accelerations, Newton imagined a bucket
swung on the end of a rope and compared it with a bucket at rest.
If the bucket is at rest, the surface of the water remains flat.
If the bucket rotates, the water surface becomes concave.
Newton argued that the concave shape of the water could not be due to
its relative motion with respect to the bucket,
and that it  proved the existence of an absolute centrifugal
acceleration.
Mach considered the same problem and criticised Newton's
reasoning \citep{Barb95}: what we observe is that the
bucket rotates with respect to the fixed stars, but, who is to say
whether the bucket is really rotating with respect to the stars at rest
or whether the stars are really rotating with respect to the bucket?
According to Newton, we should observe the concavity
in the first case and not in the second.
 This is because Newton assumed an absolute frame related to the
global
distribution of matter in the universe (the fixed stars).
Mach denied the concept of absolute acceleration.
According to him, a rotating body in a non-rotating universe
or a non-rotating body in a rotating universe should give the same result:
centrifugal force. Mach concluded that the inertial mass of a body
should  result from the contributions of all the masses in the universe.
But an obvious divergence difficulty arose, since a
homogeneous Newtonian universe with non--zero density has an infinite
mass.
Hence, Mach supported the idea of a finite universe in order to have a
finite local inertia.

By the end of the $XIX^{th}$ century, mathematicians such as Riemann had
discovered a
variety of
finite spaces without boundaries, such as the hypersphere and the
projective space.  Schwarzschild \citet{Schwa00} brought this work
to the attention of the astronomical community in 1900, and he
considered the possibility that the space could have a non--trivial
topology: ``{\it One could
imagine that as a result of enormously extended astronomical
experience,
 the entire universe consists of countless identical copies of our
Milky Way,
 that the infinite space can be partitioned into cubes each containing
  an exactly identical copy of our Milky Way.  Would we really cling
on  to the assumption of infinitely many identical repetitions of the
same world? \ldots We would be much happier with the view that these
repetitions
are illusory, that in reality space has peculiar connection properties so that
if  we leave any one cube through a side, then we immediately
reenter it    through the opposite side}" (English translation
\citealt{Schw98}).

In modern terms, Schwarzschild  told us that if the galaxies
were found
 to lie in a rectangular lattice, with images of the same galaxy
repeating
 at equivalent lattice points, then we could conclude that the Universe
is a 3--torus.

Cosmology developed rapidly after Einstein's 1915 discovery of general
 relativity.  General relativity explains gravity as the curvature of
spacetime; the latter is determined by the density of
matter--energy. The aim of relativistic cosmology is to deduce
from the gravitational field equations some physical models of
the Universe as a whole.  When Einstein (1917) assumed in his static
cosmological
solution that space was a positively--curved hypersphere, one of
his strongest motivations was to provide a model for a finite
space, although without a boundary. Following Mach and Riemann, he regarded
 the closure of space as
necessary to solve the problem of inertia and chose the hypersphere
as his model of space.

The Einstein space model cleared up most of the paradoxes
stemming from Newtonian cosmology in such an elegant way
that most cosmologists of the time adopted the new paradigm of a
closed space, without
examining other geometrical possibilities. Einstein was also
convinced that the hypershere provided not only the metric of cosmic
space -- namely its local geometrical properties -- but also its
global structure, its topology.
However, topology does not seem to have been a major preoccupation of
Einstein;
 his 1917 cosmological article did not mention any topological
alternative to the spherical space model.

 Some of his colleagues pointed out to
Einstein the arbitrariness of his choice. Indeed the  global shape of space
is not only depending
on the metric; it depends primarily on its topology, and requires a
complemenraty approach to Riemannian differential geometry.  Since
Einstein's
equations are partial derivative equations, they describe only local
geometrical properties of spacetime. The latter are contained in the
metric tensor, which enables us to calculate the components of
the curvature tensor at any non-singular point of spacetime.  But
Einstein's equations do not fix the global structure of spacetime:
to
a given metric solution of the field equations, correspond several
(and in most cases an infinite number of) topologically distinct
universe models.

Firstly, de Sitter \citet{deSitt17} noticed that the Einstein's
solution admitted a different spaceform, the 3-dimensional
projective space (also called elliptic space), obtained from the
hypersphere
by identifying antipodal points.  The projective
space has the same metric as the spherical one, but a different
topology (for instance, for the same curvature radius its volume is
half that of the spherical space).

H. Weyl also pointed out the freedom of choice between  spherical
and
elliptical topologies.  Einstein's answer  \citep{Ein1918} was
unequivocal: ``{\it Nevertheless
I have like an obscure feeling which leads me to prefer the spherical
model.
 I have a feeling that
 manifolds in which any closed curve can be continuously
contracted to a point are the simplest ones.  Other persons must
share this feeling, otherwise astronomers would have taken into
account the case where our space is Euclidean and finite.
Then the two-dimensional Euclidean space would have the connectivity
properties of a ring surface. It is an Euclidean plane in which any
phenomenon is doubly periodic, where points located in the same
periodical grid are identical. In finite Euclidean space, three
classes of non continuously contractible loops would exist.  In a
similar
 way, the elliptical space possesses a class of non continuously
contractible loops,
 contrary to the spherical case; it is the reason
why I like it less than the spherical space.  Can it be proved that
elliptical space is the only variant of spherical space?  It seems
yes
to me}".

 Einstein \citet{Ein1919} repeated his argumentation in a postcard
 sent to Felix Klein: ``{\it I would like to give you a reason why the
spherical
 case should be preferred to
the elliptical case.  In spherical space, any closed curve can be
continuously contracted to a point, but not in the elliptical space;
in other words the spherical space alone is simply-connected, not the
elliptical one [...]  Finite spaces of
arbitrary volume with the Euclidean metric element undoubtedly exist,
which can be
obtained from infinite spaces by assuming a triple periodicity,
namely identity between sets of points.  However such possibilities,
which are not taken into account by general relativity, have the
wrong
property to be multiply-connected}".

  From these remarks it seems that the Einstein's preference for
simple--connectedness of space was of an aesthetical nature, rather
than being based on physical reasoning.

  In his answer to Weyl, Einstein was
definitely wrong on the last point: in dimension three,  an infinite
number of topological variants of the spherical space -- all closed
-- do exist,
including the so-called lens spaces (whereas in dimension
two, only two spherical spaceforms exist, the ordinary sphere and
the elliptic plane).  However, nobody
knew this result in the 1920's: the topological classification of
3--dimensional spaces was still at its beginnings.  The study of
 Euclidean spaceforms started in the context of
crystallography.  Feodoroff 
\citet{Feodoroff1885} classified the 18 symmetry groups
of crystalline structures in ${\bf R}^3$, Bieberbach 
\citet{Bieberbach1911} developed
a full theory of crystallographic groups, and it took twenty years before
Novacki \citet{Novacki1934} showed how Bieberbach's results could be applied
to complete the
classification of 3--dimensional Euclidean spaceforms.  The case of
spherical spaceforms
 was first treated by Klein \citet{Klein1890} and
Killing \citet{Killing1891}.  The problem was fully solved much later
\citep{Wolf1960}.
Eventually, the classification of homogeneous hyperbolic spaces shot
forward in the 1970's; it is now an open field of intensive
mathematical research \citep{Thur79,Thur97}.

Returning to relativistic cosmology, the discovery of non-static
solutions by Friedmann \citet{Fried22} and, independently, 
Lema\^{\i}tre \citet{Lemait27},
 opened a new era for models of the universe as a whole (see, e.g.,
 \citealt{Lum97} for an epistemological analysis).
Although Friedmann and Lema\^{\i}tre are generally considered as the
discoverers
 of the big bang  concept --- at least of the notion of a dynamical
universe
evolving from an initial singularity --- one of their most original
considerations, devoted
 to the topology of space,  was
overlooked.  As they stated, the
homogeneous isotropic universe models (F--L models)  admit spherical,
Euclidean or
hyperbolic spacelike sections according to the sign of their
(constant)
curvature (respectively positive, zero or negative).
In addition, Friedmann \citet{Fried23} pointed out the topological
indeterminacy of the
solutions in his popular book on general relativity, and he
emphasized how
Einstein's theory was
unable to deal with the global structure of spacetime.  He gave the
simple example of the cylinder.  Generalizing the argument to
higher
dimensions, he concluded that several topological spaces could
be used to describe the same solution of Einstein's equations.

Topological considerations were fully developed in his second
cosmological article,
although primarily devoted to the analysis of  hyperbolic
solutions.  Friedmann \citet{Fried24} clearly outlined  the
fundamental limitations of relativistic cosmology:``{\it Without
additional assumptions, Einstein's world equations do not answer
the question of the finiteness of our universe}", he wrote.  Then he
described how space could be finite (and multi-connected) by
suitably identifying  points.  He also predicted the possible
existence of ``ghost" images of astronomical sources, since at the
same
point of a multi--connected space an object and its ghosts would
coexist.
 He  added
that ``{\it a space with positive curvature is always finite}", but he
recognized the fact that the mathematical knowledge of his time did
not allow him to ``{\it solve the question of finiteness for a
negatively--curved  space}".

 By contrast with Einstein's reasoning, it is seen that the
Russian cosmologist had no prejudice in favour of a simply-connected
topology.  Certainly, Friedmann believed that only finite volume
spaces were
physically realistic.  Prior to his discovery of hyperbolic
solutions, the
cosmological solutions derived by Einstein, de Sitter and himself
had a positive spatial curvature, thus a finite volume.  With
negatively--curved spaces,
the situation became problematic, because the
``natural" topology of hyperbolic space has an infinite volume. It is
the reason why Friedmann, in order to justify the physical pertinence
of his solutions, emphasized the possibility of compactifying
 space by suitable identifications of points.

Lema\^{\i}tre fully shared
the common belief in the closure of space. He expressed his view that
Riemannian geometry
 ``{\it dissipated the nightmare of infinite space}" \citep{Lem78}.
Thus Lema\^{\i}tre \citet{Lemait27,Lemait31} assumed
positive space curvature, but he thoroughly discussed the possibility of
projective space, that he preferred to the spherical one.
Later, he \citet{Lemait58} also noticed the possibility of
hyperbolic and Euclidean spaces with finite volumes
for describing
the physical universe.

Such fruitful ideas of cosmic topology remained widely ignored by
the main stream of big bang cosmology. Perhaps the 
Einstein--de Sitter \citet{EdS32}
model, which assumed Euclidean space and eluded the
topological question, had a  negative influence on the
development of the field.  Almost all subsequent textbooks and
monographs on relativistic cosmology assumed that the global
structure of the universe was either the finite hypersphere, or the
infinite Euclidean space, or the infinite hyperbolic space, without
mentioning at all the topological indeterminacy.  As a consequence,
some
 confusion settled down about the real meaning of the terms ``open"
and
``closed" used to characterize the F--L  solutions. Whereas they apply
correctly to
 time evolution (open models stand for ever--expanding universes,
closed
 models stand for expanding--contracting solutions), they do not
 properly describe the extension of
space (open for infinite, closed for finite). Nevertheless it is
still frequent to read that the (closed) spherical
model has a finite volume whereas the (open) Euclidean and hyperbolic
models
have infinite volumes. The
correspondance is true only in the very special case of a
simply--connected topology {\it and} a zero cosmological constant.
According to Friedmann's original remark, in order to know if a
space is finite or infinite, it is not sufficient to determine the
sign of its spatial curvature, or equivalently in a cosmological
context to measure the ratio $\Omega_{0}$
of the average density to the critical value: additional assumptions
are necessary --- those arising from topology, precisely.

The idea of a multi--connected universe has never been entirely
forgotten. In a very comprehensive article, Ellis \citet{Ell71} detailed the
classification of 3--dimensional Riemannian manifolds useful for
cosmology and started to explore the observational consequences of a
toroidal universe. Others, such as Sokoloff and Schvartsman, Fang and
Sato, Gott, Fagundes, investigated flat and non-flat spaces (see
complete references in LaLu95).
Nevertheless, until 1995, investigations in cosmic topology were
rather scarce.
  From an epistemological point of view, it
seems that the prejudice in favour of simply--connected (rather
than multi--connected) spaces was of the same kind as the prejudice
 in favour of static (rather than dynamical)
cosmologies during the 1920's.  At a first
glance, ``Occam's razor" (often useful in physical modelling)
could
be invoked to preferably select the simply--connected topologies.
However,
on the theoretical side,
new approaches to spacetime, such as quantum cosmology,
suggest that the smallest closed hyperbolic manifolds are favoured,
thus providing a new paradigm for
what is the
``simplest" manifold. On the observational side, present astronomical data
(e.g. \citealt{Dek96})
indicate that the average density of the observable universe is less than the
critical value  ($\Omega_{0} = 0.3 - 0.4$), thus suggesting that we live in a
negatively--curved F--L universe (unless
the cosmological constant is positive and large enough). Putting together
theory and observation,
cosmologists must face the fact that a negatively--curved space can have a
finite volume, and that in that case it must be multi--connected.

In the last few decades, much effort in observational and theoretical
cosmology
has been directed towards determining the curvature of
the universe. The problem of the topology of spacetime was generally
ignored within the framework of classical relativistic cosmology. It
began to be taken seriously discussed in quantum gravity for various
reasons: the spontaneous birth
of the Universe from a quantum vacuum  requires the Universe
to have compact spacelike hypersurfaces, and the closure of space is
 a necessary condition to render tractable the integrals of quantum
gravity \citep{Atk82}.

Further work in quantum gravity and particle physics
which explores ideas of space-time
topology globally and/or as the basis of a quantum gravity
theory includes that of
\citet{Hawking84,ZelG84,MadS97,Carl98,Ion98,DowS98,DowG98,Rosal98,eCF98,Spaa99}.

However, the topology of spacetime also enters in a
fundamental way in classical general relativity.  Many
cosmologists were surprisingly unaware of how topology and cosmology
could fit together to provide new insights in universe models.  Aimed
to create a new
interest in the field of cosmic topology, the
 review by LaLu95 stressed what multi--connectedness of
the
Universe would mean and on its observational consequences. However
two different
papers (\citet{Stev93} ; \citet{deOliv95})
declared that the small
universe idea was ``no longer an interesting cosmological model".
However, the authors drew very general conclusions from
considering only a few special 3-manifolds
in the Euclidean case, and by adopting assumptions on the
primordial perturbation spectrum which are observationally justified on
large scales only by assuming simple connectedness.
They did not take into account a very interesting class
of realistic universe models, namely the compact
hyperbolic
manifolds, which require a quite different treatment. Ironically,
approximately the same amount of
papers in cosmic topology have been
published within the last 3 years as in the previous 80 years !

\section{Mathematical background}

\subsection{Basics of Topology}

The topological properties of a manifold are  those which
remain insensitive to continuous transformations. Thus, size and
distance are in some sense ignored in topology:  stretching,
squeezing
or ``kneading'' a manifold change the metric but not the topology;
cutting, tearing or making holes and handles change the latter.
It is often possible to visualize two--dimensional manifolds by
representing them as embedded in three--dimensional Euclidean
space (such a  mapping does not necessarily
exist however, e.g. the Klein bottle and the flat torus).
Three--dimensional manifolds require the
introduction of more abstract representations, using, for example, the
{\sl
fundamental domain}.

\begin{figure}
\centerline{\epsfxsize=8cm
\zzz{\epsfbox[145 305 419 510]{"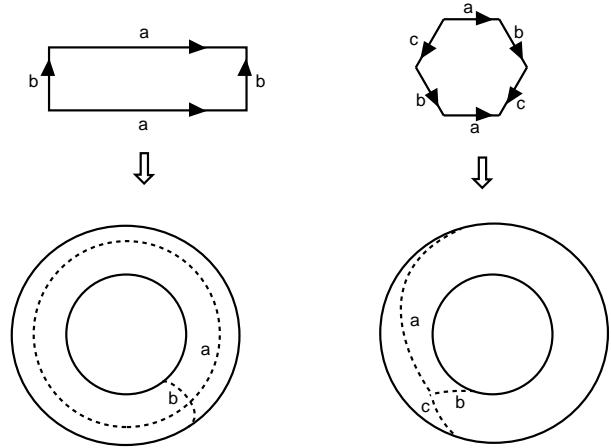"}}
{\epsfbox[145 305 419 510]{"`gunzip -c torus.ps.gz"}}
}
\caption{Construction of the flat torus from a rectangle and from an
hexagon with suitable identifications}
\label{torus}
\end{figure}

The different topological
surfaces can be represented by polygons
whose edges are suitably identified by pairs. As we have seen in the
introduction, identifying one pair of
opposite
edges of a square gives a portion of a cylinder; then, stretching the
cylinder and bending it around in such a way that one
can glue together the two circular ends generates a simple
torus, a
{\sl closed} surface (figure~\ref{torus}). The torus is thus
topologically equivalent to a rectangle
with opposite edges identified. The rectangle is
called a {\sl fundamental domain} of the torus.  From a topological
point of view (namely without reference to  size),
the fundamental domain can be chosen in different ways: a square,
a rectangle, a parallelogram,  even a hexagon
(since the plane can be tiled by hexagons, the flat torus
can be also represented by a hexagon with suitable
identification of edges).

\begin{figure}
\centerline{\epsfxsize=6cm
\zzz{\epsfbox[185 502 295 621]{"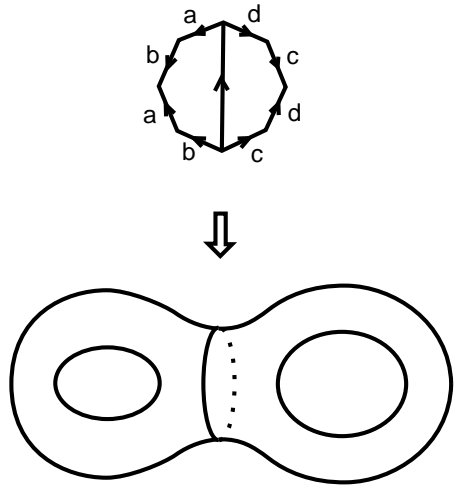"}}
{\epsfbox[185 502 295 621]{"`gunzip -c gtorus.ps.gz"}}
}
\caption{The 2-torus as the connected sum of two
simple tori}
\label{gtorus}
\end{figure}

The gluing method described above becomes extremely fruitful
when the surfaces are more complicated. A two--dimensional g--torus
$T_g$ is a torus with $g$ holes.   $T_g$ can be constructed as
the {\sl connected sum} of $g$ simple tori (figure
\ref{gtorus}). The g--torus is therefore topologically equivalent to a
connected sum of $g$ squares whose opposite edges have been
identified. This sum is itself topologically equivalent to a 4g--gon
where all the vertices are identical with each other and the sides are
suitably identified by pairs.

It would be tempting to visualize the g--torus by gluing together
equivalent  edges, like for the simple torus. But such an operation
is not
straightforward when $g \ge 2$. All the vertices
of the polygon correspond to the same point of the surface. Since
the polygon has at least 8 edges, it is necessary to make the
internal angles thinner in order to fit them suitably around a single
vertex. This can only be achieved if the polygon is represented in the
{\sl hyperbolic plane} ${\bf H}^2$ instead of the Euclidean plane ${\bf R}^2$:
this increases the area and decreases the angles. The more
angles to fit together, the thinner they have to be and the greater the
surface. The
g-torus $(g \geq 2)$ is therefore a {\sl compact surface of negative
curvature}.

When one deals with more than two dimensions, the gluing method
remains the simplest way to visualize spaces. By analogy with the
two-dimensional case, the three-dimensional simple torus (also
referred to as the {\sl hypertorus}) is
obtained by identifying the opposite faces of a parallelepiped. The
resulting
 volume is finite. Let us imagine a light source at our position,
immersed in such a structure. Light emitted backwards crosses the
face
of the parallelepiped behind us and reappears on the opposite face in
front  of
us; therefore, looking forward we can see our back (as in the
spherical
universe model of Einstein). Similarly,  we
see in our right our left profile, or upwards the  bottom of
our feet. In fact, for light emitted isotropically, and for an
arbitrarily large time to wait, we could observe ghost images
of any object viewed arbitrarily close to any angle. The
resulting visual effect would be comparable (although not identical)
to
what  could be seen from inside a parallelepiped of which the internal faces
are covered with mirrors. Thus one would have the visual
impression  of infinite space, although the real space is closed.
More generally, any three dimensional compact manifold
can be represented as a polyhedron -- what we define later
more precisely as the fundamental domain (hereafter $FD$) -- whose
faces  are suitably identified by pairs. But, as soon as the number
of faces of the $FD$ exceeds 8, the compact manifold resulting from
identifications  cannot be developed into the Euclidean space
${\bf R}^3$: the
$FD$ must be built  in hyperbolic space ${\bf H}^3$
 in order to fit all the angles together at the vertices.

The general method for classifying the topologies of a given manifold $\cal
M$  is:
\begin{itemize}
\item to determine its universal covering space $\widetilde{ \cal
M}$
\item  to find the fundamental domain $FD$
\item  to calculate the holonomy group acting on the $FD$.
\end{itemize}

 All these concepts have very  formal and abstract
definitions that can be found in classical textbooks in topology. We
introduce below basic definitions.

The strategy for characterizing spaces is to
produce invariants which capture the key features of the topology and
uniquely specify each equivalence class. The topological invariants
can take
many forms. They can be just numbers, such as the dimension of the
manifold, the degree of connectedness or the
Poincar\'e -- Euler characteristic. They can also be whole
mathematical structures, such as the homotopy groups.

 A loop at ${\bf x} \in
\cal{M}$  is any path which starts at $\bf x$ and ends at $\bf x$.
Two loops
 $\gamma$ and $\gamma ' $ are homotopic if $\gamma$ can be
continuously
deformed into  $\gamma '$.
The manifold $\cal{M}$ is {\sl simply--connected} if, for any $\bf x$,
two any  loops through $\bf x$ are homotopic. Equivalently,
it is {\sl simply--connected} if
every loop
is  homotopic to a point.  If not, the manifold is said to be {\sl
multi--connected}.
Obviously, the Euclidean spaces ${\bf R}^1$, ${\bf R}^2$,\ldots, ${\bf
R}^n$,
and the  spheres ${\bf S}^2$,~ ${\bf S}^3$,\ldots, ${\bf S}^n$ are
simply--connected, whereas the circle ${\bf S}^1$, the  cylinder
${\bf S}^1 \times {\bf R}$ and the torus ${\bf S}^1 \times {\bf S}^1$ are
multi--connected.

The study of homotopic loops in a manifold $\cal M$ is a way of
detecting holes or handles. Moreover the equivalence classes of
homotopic loops can be endowed with a group structure, essentially
because loops can be added by joining them end to end.  The group of
loops is called the first
homotopy group at $\bf x$ or, in the terminology originally
introduced by Poincar\'e, the {\sl fundamental group}
$\pi _1($$\cal M$,${\bf x})$. The fundamental group is independent of
the base point: it is a topological invariant of the manifold.

\begin{figure}
\centerline{\epsfxsize=4cm
\zzz{\epsfbox[235 510 336 572]{"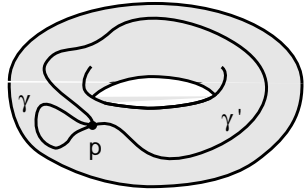"}}
{\epsfbox[235 510 336 572]{"`gunzip -c homotopy.ps.gz"}}
}
\caption{ Classes of homotopy of a
torus $\protect{\bf S}^1  \times {\bf S}^1$.
Loops can wind m times around the central hole and n times
around the body of the torus. Thus the fundamental group consists
of pairs $(m,n)$ of integers with addition $(m,n) + (p,q) = (m+p,
n+q)$. In other words it is isomorphic to $\protect{\bf Z}\protect
\oplus\protect {\bf Z}$.}
\label{homotopy}
\end{figure}

For surfaces, multi--connectedness
means that the fundamental group is non trivial: there is at least
one loop that cannot be shrunk to a point.
But in higher dimensions the problem is more complex because
loops, being only one--dimensional structures, are not sufficient to
capture all the topological features of the manifolds. The purpose of
algebraic topology, extensively developed during the twentieth
century, is to generalise the concept of homotopic loops and to
define higher homotopy groups. However the fundamental group
(the first homotopy group) remains essential.

To define the universal covering space, start with a manifold $\cal
M$ with metric $\bf g$. Choose a base
point {\bf x} in $\cal M$ and consider the differents paths from {\bf
x} to another point {\bf y}. Each path belongs to a homotopy class
$\gamma$ of loops at {\bf x}. We construct the universal covering
space as the new manifold  ($\widetilde{ \cal M}$,$\tilde{\bf g}$)
such
that each point $\tilde{\bf y}$ of $\widetilde{ \cal M}$ is obtained
as a
pair (${\bf y},\gamma$), {\bf y} varying over the whole of $\cal M$
while {\bf x} remains fixed and $\gamma$ varies over all homotopy
classes. The metric $\tilde{\bf g}$ is obtained
by
defining the interval from $\tilde{\bf x} = ({\bf x},\gamma)$ to a
nearby point $\tilde{\bf x'}= ({\bf x'},\gamma)$ in $\widetilde{ \cal
M}$ to be equal to the interval from $\bf x$ to $\bf x'$ in $\cal M$.
By construction, ($\widetilde{ \cal
M}$,$\tilde{\bf g}$) is locally indistinguishable from (${ \cal
M}$,${\bf g}$). But its global -- namely topological -- properties
can be quite different. It is clear that,
 when $\cal M$ is simply--connected, it is  identical to its universal
covering space $\widetilde{ \cal M}$. When $\cal M$ is
multi--connected, each point of $\cal M$ generates an infinite number
of
points in $\widetilde{ \cal M}$. The universal covering space can thus
be thought of as an ``unwrapping" of the original manifold.

Consider a point ${\bf x}$ and a loop $\gamma$ at {\bf x} in $\cal M$.
If $\gamma$ lies entirely  in a simply-connected domain of
$\cal M$, (${\bf x},\gamma$) generates a single point $\tilde{\bf x}$
in $\widetilde{ \cal M}$.
Otherwise, it generates additional points $\tilde{\bf x'},\tilde{\bf
x''}
$, \ldots which are said to be {\sl homologous} to $\tilde{\bf x}$.
The displacements $\tilde{\bf x} \mapsto \tilde{\bf x'}$, $\tilde{\bf
x} \mapsto \tilde{\bf x''}$, \ldots  are isometries and form the
so-called
{\sl holonomy group} $\Gamma$ in $\widetilde{ \cal M}$. This group
is discontinuous, i.e., there is a non zero  shortest distance
between any
two homologous points, and the generators of the group (except the
identity) have no fixed point. This last property is very restrictive
(it
excludes for instance the  rotations) and allows the classification of all
possible holonomy groups.

Equipped with such properties, the holonomy group is said to
act freely and discontinuously on $\widetilde{ \cal M}$. The holonomy
group is isomorphic
to the fundamental group $\pi_1(\widetilde{\cal M})$.

The geometrical properties of a manifold $\cal M$  within a
simply--connected domain are the same as those of its development in
the universal covering $\widetilde{\cal M}$.  The largest
simply--connected domain containing a given point
$\bf x$ of $\cal M$, namely the set $ \bigl\{ {\bf y} \in {\cal M},
d(\tilde{\bf y},\tilde{\bf x})
\le d(\tilde{\bf y},\gamma (\tilde{\bf x})),  \forall \gamma \in \Gamma
\bigr\}$, is called the
{\sl fundamental domain} ($FD$).

The $FD$ is always convex and has a finite number of faces (due
to the fact that the holonomy group  is discrete). These faces are
homologous by pairs ; the
displacements carrying one face to another are the generators
of the holonomy group $\Gamma$.

 In two dimensions, the $FD$ is a surface whose boundary is
constituted by lines, thus a polygon. In three dimensions, it is
a volume bounded by  faces, thus a polyhedron.

 The configuration formed by the fundamental polyhedron $FD$ and its
images $\gamma $$FD$ ($\gamma \in \Gamma$) is called a
{\sl tesselation} of $\widetilde{ \cal
M}$, each image $\gamma $$FD$ being a cell of the tesselation.

The $FD$ has two major advantages:
\begin{itemize}
\item The fundamental group of a given topological
manifold $\cal M$ is isomorphic to the fundamental group of the
$FD$. Routine methods are available to determine
the holonomy group of a polyhedron.
\item The $FD$ allows one to  represent any curve in
$\cal M$, since any portion of a curve lying  outside the $FD$ can
be carried inside it by appropriate holonomies.
\end{itemize}

\subsection{Two-dimensional manifolds}

In addition to pedagogical and illustrative interest, the
classification of two--dimensional Riemannian surfaces plays an
important role in physics for understanding (2+1)--dimensional
gravity, a toy model to gain insight into the real world of
(3+1)--dimensional  quantum gravity.

\begin{figure}
\centerline{\epsfxsize=10cm
\zzz{\epsfbox[116 400 427 622]{"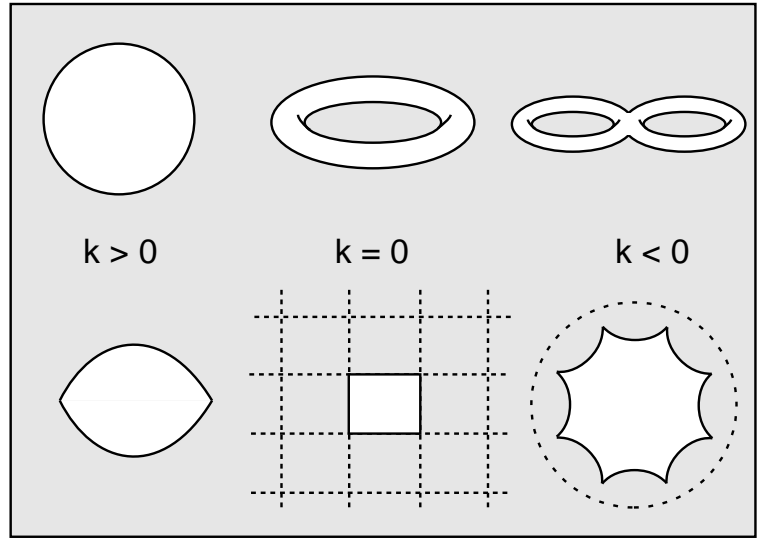"}}
{\epsfbox[116 400 427 622]{"`gunzip -c univcov.ps.gz"}}
}
\caption{ The three kinds of geometries for
Riemannian surfaces, as closed surfaces ``squeezed'' into ${\bf R}^3$,
and as fundamental domains ($FD$'s). The covering space for $k > 0$
is the sphere ${\bf S}^2$, and the covering spaces for the other
two curvatures are shown schematically.
\label{univcov}}
\end{figure}

Any Riemannian surface is
homeomorphic to a surface admitting a metric with constant curvature
k. Thus, any Riemannian surface can be expressed as the quotient
  $\cal{M}$ $= \widetilde{\cal M}/\Gamma$,
 where the universal covering space $\widetilde{\cal M}$ is
either (figure \ref{univcov}):
 \begin{itemize}
 \item the Euclidean plane ${\bf R}^2$  if $k = 0$
 \item the sphere ${\bf S}^2$ if $k > 0$
\item  the hyperbolic plane ${\bf H}^2$  if $k < 0$.
\end{itemize}
and $\Gamma$ is a discrete subgroup of isometries  of
$\widetilde{\cal M}$ without fixed point.

To characterise the quotient spaces we adopt the following
abbreviations:

 C = closed, O = open, SC = simply--connected, MC =
multi--connected, OR = orientable, NOR = non-orientable.

\begin{figure}
\centerline{\epsfxsize=10cm
\zzz{\epsfbox[134 364 451 686]{"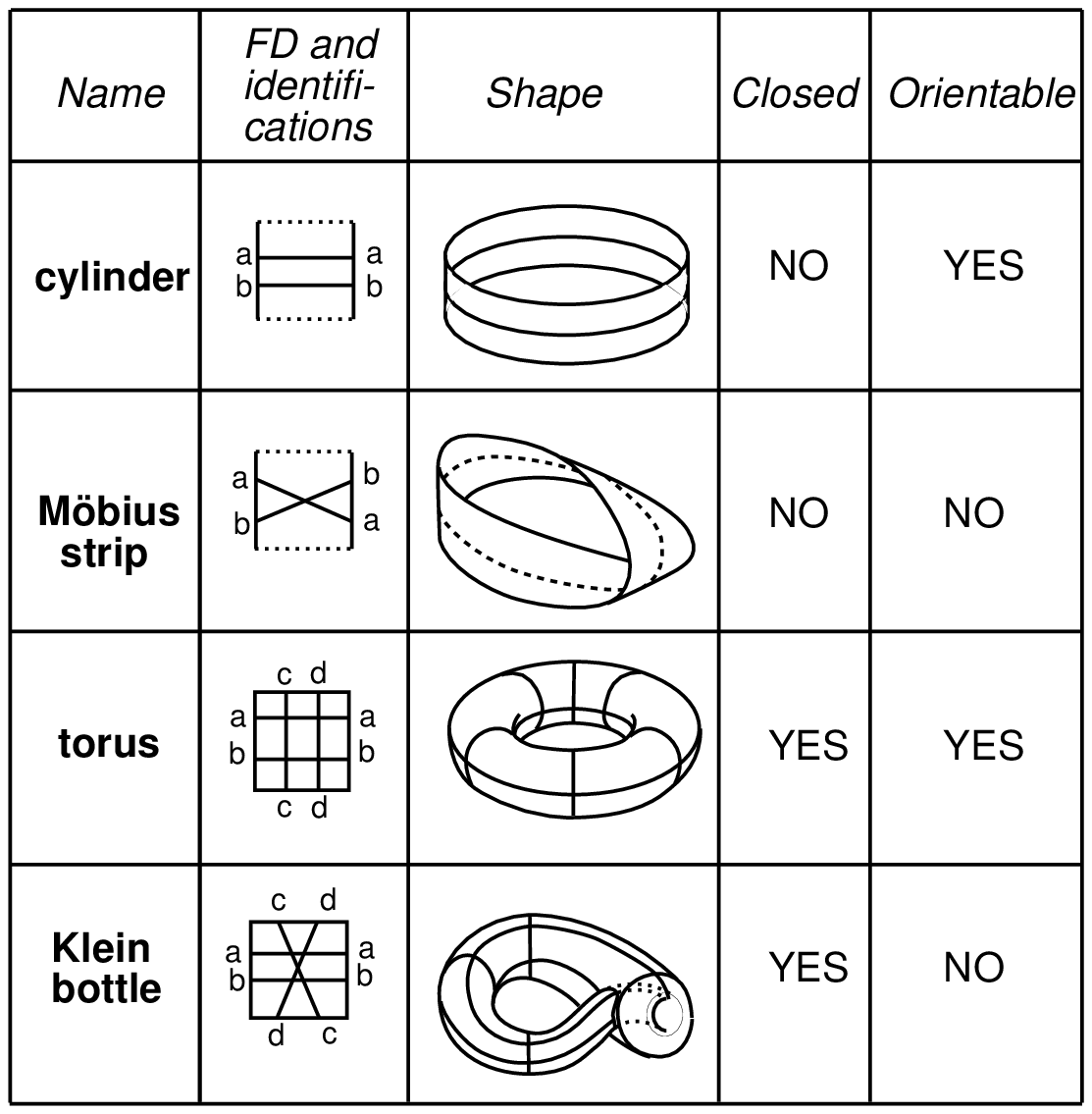"}}
{\epsfbox[134 364 451 686]{"`gunzip -c 2surfaces.ps.gz"}}
}
\caption{ The four types of multi-connected Euclidean surfaces}
\label{2surfaces}
\end{figure}

Locally Euclidean surfaces fall into only 5 types: the
simply--connected Euclidean plane itself ${\bf R}$,  the
multi--connected
cylinder ${\bf R}\times {\bf S}^1$, the M\"obius band,  the torus
${\bf S}^1 \times {\bf S}^1$ and the Klein bottle. Their characteristics are
summarized
in figure \ref{2surfaces}. The M\"obius band and the Klein
bottle are not  orientable. The torus and the Klein bottle are closed
spaces.

There are only two spherical surface forms (this result has been
generalized to any constant positive
curvature manifold of {\it even} dimension):
 \begin{itemize}
 \item the sphere ${\bf S}^2$ itself: C, SC, OR
 \item the projective plane ${\bf P}^2 \equiv {\bf S}^2/{\bf Z}_2$ (also
called
the elliptic plane): C, MC, NOR.
\end{itemize}

Whereas the surface of the unit sphere is $4\pi$, the surface of the
unit
projective plane is only $2\pi$, and its diameter, i.e., the distance
between the most widely separated points, only $\pi/2$.

The hyperbolic plane
${\bf H}^2$, historically known as the Lobachevski space, is
difficult to visualize because it cannot be isometrically imbedded in
${\bf R}^3$. Nevertheless it can be thought of as a surface with a
saddle point
at every point.

 The full
isometry group of ${\bf H}^2$ is $PSL(2,{\bf R}) \equiv
SL(2,{\bf R})/{\bf Z}_2$, where $SL(2,{\bf R})$ is the group of real $2
\times 2$
matrices with unit determinant.

\begin{figure}
\centerline{\epsfxsize=8cm
\zzz{\epsfbox[203 322 345 464]{"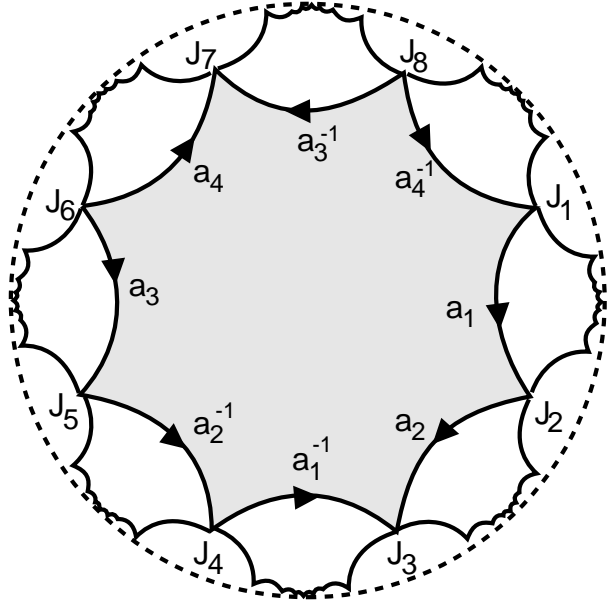"}}
{\epsfbox[203 322 345 464]{"`gunzip -c escher.ps.gz"}}
}
\caption{ Tesselation of ${\bf H}^2$ by octagons in the Poincar\'e
representation. The vertices have the coordinates $z(J_k) =
0.841 ~exp~\frac{(3-2k)\pi i}{8}$}
\label{escher}
\end{figure}

 The  best known example of a compact hyperbolic
surface is the 2-torus $T^2$.  In  this case, the
$FD$ is a regular octagon with pairs of sides identified.  In the
Poincar\'e representation of ${\bf H}^2$, the $FD$ appears
curvilinear.  The pavement of the unit disk by homologous octagons
(which appear distorted in this representation)  corresponds to the
tesselation of ${\bf H}^2$ by regular octagons (figure \ref{escher}).
The famous Dutch artist M.--C. Escher has designed
fascinating  drawings and prints using such tilings of the hyperbolic
plane.

The topological classification of locally hyperbolic surfaces is
complete only for the {\sl compact} ${\bf H}^2/ \Gamma$, which
fall into one of the following categories:

\begin{itemize}

\item g--torus $T_g$, $g \ge 2$ (connected sum of g simple tori):
C,
MC, OR

\item connected sum of n projective planes: C, MC, NOR

\item connected sum of a compact orientable surface (${\bf S}^2$ or
$T_g$) and
of a projective plane or  a Klein bottle: C, MC, NOR
\end{itemize}

All of these surfaces have a finite area bounded below by $2\pi$, and
a
diameter greater than cosh$^{-1}(4) \approx 2.06$.

There are an infinite number of {\sl non--compact} locally
hyperbolic surfaces. Clearly,``almost all" Riemannian surfaces are
hyperbolic, since:

  - Any open surface other than the Euclidean plane, the cylinder and
the
M\"obius band is homeomorphic to a locally hyperbolic surface, for
example an
hyperbolic plane with or without handles.

- Any closed surface which is not the sphere, the projective plane,
the torus
or the Klein bottle is homeomorphic to a locally hyperbolic surface.

\subsection{Three-dimensional manifolds of constant curvature}

Any three-dimensional Riemannian
manifold $\cal M$ admitting at least a 3--dimensional discrete
isometry group $\Gamma$ simply transitive on $\cal M$ is locally
homogeneous
and can be written as the quotient
$\widetilde{\cal M} / \Gamma$, where  $\widetilde{ \cal M}$ is the
universal covering space of $\cal M$. Let  $G$ be the full group
of isometries of $\cal M$ (containing $\Gamma$ as a discrete
subgroup).
In the terminology used in the theory of classification of compact
three--manifolds, $\cal M$ is said to admit a {\sl geometric
structure}
modelled on $(\widetilde {\cal M}, G)$.

Thurston has classified the homogeneous
three--dimensional geometries into eight distinct types, generally
used by mathematicians.

On the other hand, the Bianchi types are defined  from the
classification of all simply-transitive 3--dimensional Lie
groups.  Since the
 isometries of a Riemannian manifold form a Lie group, the Bianchi
classification is used  by workers in relativity and cosmology for the
description of spatially homogeneous spacetimes.
The correspondance between the locally homogeneous 3--geometries
in Thurston's sense and the Bianchi--Kantowski--Sachs classification
 of homogeneous cosmological models has recently been fully clarified
 by \citet{Rainer96}.

 Cosmology, however, focuses mainly on
locally homogeneous and isotropic spaces, namely those admitting
one of the 3 geometries of constant curvature.
Any compact 3--manifold $\cal{M}$ with constant curvature $k$ can be
expressed as the quotient
${\cal M}~\equiv ~\widetilde{\cal M}/\Gamma$,
 where the universal covering space $\widetilde{\cal M}$ is either:
\begin{itemize}
\item the 3-sphere ${\bf S}^3$ if $k > 0$
\item  the Euclidean space ${\bf R}^3$ if $k = 0$
\item the hyperbolic 3-space ${\bf H}^3$ if $k < 0$.
\end{itemize}
and $\Gamma$ is a subgroup of isometries of $\widetilde{\cal M}$
acting
freely and discontinuously.

We give below a schematic description of such spaces.

\subsubsection {Spherical space forms}\label{positivem}

 Three--manifolds of constant positive curvature have
${\bf S}^3$, which is compact,  as universal
 covering space. As a consequence they are all compact.

The  metric on ${\bf S}^3$ may be written as

\begin{equation}\label{ms3}
d\sigma ^2  =
 R^2 \{ d\chi ^2 + sin^2\chi (d\theta ^2 + sin^2 \theta d\phi ^2)\}.
\end{equation}

The volume is $2 \pi ^2 R^3$.

The full isometry group of  ${\bf S}^3$ is $SO(4)$.
The admissible subgroups $\Gamma$ of $SO(4)$ without fixed point,
acting freely and discontinuosly on ${\bf S}^3$, are:

\begin{itemize}
\item the cyclic groups of order p, $Z_p$  ($p \ge 2$).
\item the dihedral groups of order $2m$, $D_m$ ($m > 2$).
\item the polyhedral groups, namely:

\begin{itemize}

 \item the group $ T$  of the tetrahedron (4 vertices, 6 edges, 4
faces), of order 12;

\item the group  $O$  of  the octahedron (6 vertices, 12 edges, 8
faces), of order 24 ;

 \item the group  $I$  of the icosahedron (12 vertices, 30 edges, 20
faces), of order 60.
\end{itemize}
 \end{itemize}

All
the homogeneous spaces of constant positive curvature are
obtained by quotienting ${\bf S}^3$ with the groups described above.
 They are in infinite number due to parameters $p$ and $m$.

The volume of  $\cal M$$ = {\bf S}^3 /\Gamma$ is simply
\begin{equation}\label{vol}
vol({\cal M}) = 2 \pi ^2~R^3 / \mid \Gamma \mid
\end{equation}
 where $\mid \Gamma \mid$ is the order of the group $\Gamma$.
For topologically complicated spherical 3--manifolds,
$\mid \Gamma \mid$ becomes large and $vol(\cal{M})$ is small. There
is no lower bound since $\Gamma$ can have an arbitrarily large number
of elements. Hence
\begin{equation}\label{vol1}
0 < vol({\cal M}) \leq 2 \pi ^2~R^3
\end{equation}
In contrast, the {\sl diameter}, i.e., the maximum distance
between two points in the space, is bounded below by $\frac{1}{2}
arccos((1/\sqrt3)~cot(\pi/5)~R \approx 0.326~R$, corresponding to a
dodecahedral space.

As well--known examples one can cite

\begin{figure}
\centerline{\epsfxsize=5cm
\zzz{\epsfbox[202 268 322 421]{"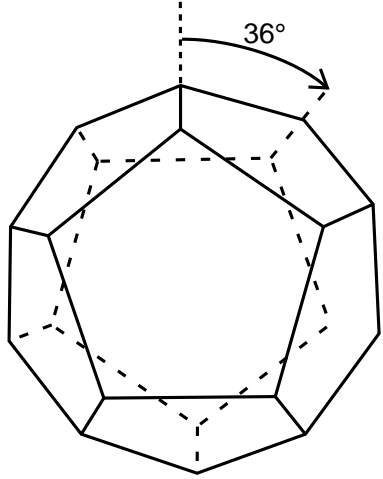"}}
{\epsfbox[202 268 322 421]{"`gunzip -c poincare.ps.gz"}}
}
\caption{ The fundamental domain of the Poincar\'e spherical space}
\label{poincare}
\end{figure}

\begin{itemize}
\item The projective space
 ${\bf P}^3 = {\bf S}^3/{\bf Z}_2$, obtained by identifying
diametrically opposite points on ${\bf S}^3$. It was used by de Sitter
and Lema\^{\i}tre as the space structure of their cosmological models.

\item The lens spaces $ {\bf S}^3/{\bf Z}_p$. The
simplest one,  $ {\bf S}^3/{\bf Z}_3$,divides
${\bf S}^3$
 into  6 fundamental cells, each having a lens form.

\item  The Poincar\'e dodecahedral space is an example of
${\bf S}^3/I$. The fundamental
polyhedron is a regular dodecahedron of which the faces are pentagons.
The compact  space is obtained by identifying the opposite faces after
rotating by  $1/10^{th}$ turn in the clockwise direction around the
axis orthogonal to the
 face (figure \ref{poincare}). This configuration involves 120 successive
operations and
 gives already some idea of the extreme complication of such
multi--connected topologies.
\end{itemize}

\subsubsection{Euclidean space forms}
\label{flatm}

The line element for the universal covering space ${\bf R}^3$ may be
written
as:
\begin{equation}\label{e3m}
d\sigma ^2 = R^2 \{ d\chi ^2 + \chi^2 (d\theta ^2 + sin^2 \theta
d\phi ^2)\}
\end{equation}

Its full isometry group is $G = ISO(3) \equiv {\bf R}^3 \times SO(3)$,
and
the generators of the possible holonomy groups $\Gamma$ (i.e.,
discrete
subgroups without fixed point) include
 the identity, the translations, the glide reflections and the
helicoidal
motions
occurring in various combinations.  They generate  18 distinct types
of locally Euclidean spaces. The 17
multi--connected space forms are in correspondance with the 17
crystallographic groups discovered more than a century ago by
Feodoroff. Eight forms are open (non compact), ten are closed
(compact).

The compact models can be better visualised by identifying
appropriate  faces of fundamental polyhedra. Six of them are
orientable.
The fundamental polyhedron can be a parallelepiped or a hexagonal
prism, with various possible possible identifications. The 3-torus is
the simplest one.
Their volume is arbitrary, since there is no rigidity theorem linking
the curvature radius to the topological lengths. Until 1985
\citep{Fag85,Fag89,Fag96},
such spaces were nearly the only ones to be
explored for discussion of cosmic topology.

\begin{figure}
\centerline{\epsfxsize=10cm
\zzz{\epsfbox[156 319 425 430]{"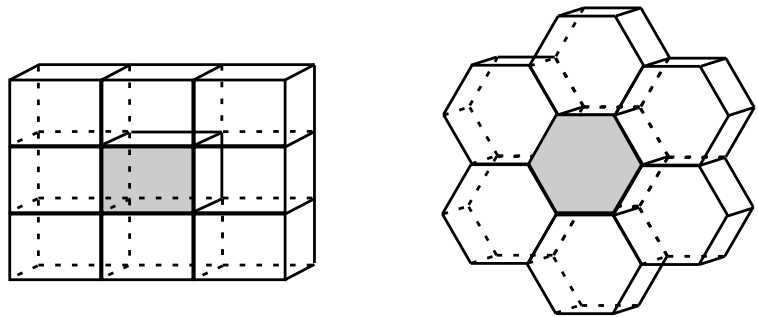"}}
{\epsfbox[156 319 425 430]{"`gunzip -c 3euclidien.ps.gz"}}
}
\caption{ Tesselation of ${\bf R}^3$ by parallelepipeds or hexagonal
cells}
\label{3euclidien}
\end{figure}

\subsubsection {Hyperbolic space forms}
\label{negativem}

Locally hyperbolic manifolds are less well understood than the
other homogeneous spaces.  However, according to the
pioneering work of Thurston, ``almost all''
3--manifolds can be endowed with a hyperbolic structure.

The induced metric on ${\bf H}^3$ may be written as
\begin{equation} \label{h3m}
d\sigma ^2 =
R^2~ \Bigl\{~ d\chi ^2 + sinh^2 \chi ~(d\theta^2 + 
sin^2 \theta~d\phi ^2)
\Bigr\}
 \end{equation}

The volume of ${\bf H}^3$ is infinite. Its  group of isometries is
isomorphic to
$PSL(2,{\bf C})$, namely the group of
 fractional linear transformations acting on the complex plane.
Finite subgroups  are discussed in \citet{Bea83}.

 In hyperbolic geometry there is an essential difference between the
2--dimensional case  and higher dimensions. A {\sl surface} of genus
$g
\ge 2$ supports uncountably many non equivalent hyperbolic metrics.
But for $n \ge 3$, a connected oriented n--manifold supports {\it at
most
one} hyperbolic metric. More precisely,  the {\sl rigidity theorem}
proves that (for a fixed value of $R$)
 if two hyperbolic  manifolds, with dimension $n \ge 3$,
have
isomorphic fundamental groups, they are necessarily isometric to each
other. It follows that, for $n \ge 3$, the volume of a manifold and the
lengths of its closed geodesics are topological invariants.
  This suggested the idea of  using
the volumes to classify the topologies, which could have seemed, at a
first glance, contradictory with the very purpose of topology.

Each  type of topology is  characterized by some lengths. For compact
locally Euclidean  spaces, the fundamental domain may
possess arbitrary volume, but no
more than eight faces. In the spherical case, the volume of ${\bf S}^3/
\Gamma$ is finite and  is that of ${\bf S}^3$,
the maximum possible value, divided by a whole number.
By contrast,  it is possible to
tesselate ${\bf H}^3$  with polyhedra having an arbitrarily large
number of
faces. In  the three-dimensional hyperbolic case, the possible
values for the   volume of the $FD$ are bounded from below.   In
other words, there exist hyperbolic 3--manifolds with volumes
arbitrarily close to a minimal volume for all hyperbolic 3--manifolds.

Particular interest has been taken by various authors in computing the
volumes of compact hyperbolic manifolds.
 Each topology has a specific volume measured in curvature radius
units.
The absolute lower bound for the volume of CHMs is given by $V_{min}
= 0.16668$ \citep{Gabai96}. However, it may have little effect on
cosmological applications.
The reason is that the true lower bound is expected to be 0.942707,
corresponding to the smallest CHM that is presently
known
\citep{Weeks85,Matv88}. The new $V_{min}$ bound
represents
an improvement in the techniques
of the proof, not an increase in the expected size of the smallest
hyperbolic manifold.

In view of cosmological observational effects, the smaller the value
of $vol_{min}$, the more interesting the
corresponding manifold for cosmology. Thus the  Weeks space has been
specially studied, by \citet{Fag93} and more recently by
\citet{LLU98}. Its $FD$ is a polyhedron with 26
vertices and 18 faces, of which 12 are pentagons and 6 are
quadrilaterals. Its outer structure
is represented in figure (\ref{weeks}), the Klein coordinates of the
vertices and the
18 matrix representations of the generators of the holonomy group are
given in \citet{LLU98}.

\fweeks

The Weeks manifold leaves room for many topological lens effects,
since the volume of the observable universe is about 200 times larger
than the volume of
Weeks space for $\Omega_{0} = 0.3$. Indeed,
many CHMs have
geodesics shorter than the curvature radius, leaving room to fit a
great many
copies of a fundamental polyhedron within the horizon radius, even
for manifolds of volume $\sim$ 10.
The publicly available program {\sc SnapPea} (Weeks) is
specially useful to unveil the rich structure of CHMs. Several
millions of CHMs with volume less than 10 could be calculated.
Table~\ref{t-snappea}
summarizes a sample of the results ($r_{-}$ is the radius of
the largest
sphere in the covering space  which can be inscribed in the
fundamental polyhedron,
$r_{+}$ is the radius of the smallest
sphere in the covering space in which the fundamental polyhedron can
be inscribed, $l_{min}\equiv 2\rinj$ is the
length of the shortest geodesic; see Fig.~\ref{f-rprmrinj}).

\begin{table}
\caption{{Small CHMs from SnapPea} \label{t-snappea}}
\begin{center}
\begin{tabular}{ccccc}
\hline
Name & Volume& $r_-$ & $r_+$ & $l_{min}$ \\
\hline
WMF   & 0.9427 & 0.5192 & 0.7525 & 0.5846 \\
Thurston   & 0.9814 & 0.5354 & 0.7485 & 0.5780 \\
s556(-1,1) & 1.0156 & 0.5276 & 0.7518 & 0.8317 \\
m006(-1,2) & 1.2637 & 0.5502 & 0.8373 & 0.5750 \\
m188(-1,1) & 1.2845 & 0.5335 & 0.9002 & 0.4804 \\
v2030(1,1) & 1.3956 & 0.5483 & 1.0361 & 0.3662 \\
m015(4,1)  & 1.4124 & 0.5584 & 0.8941 & 0.7942 \\
s718(1,1)  & 2.2726 & 0.6837 & 0.9692 & 0.3392 \\
m120(-6,1) & 3.1411 & 0.7269 & 1.2252 & 0.3140 \\
s654(-3,1) & 4.0855 & 0.7834 & 1.1918 & 0.3118  \\
v2833(2,3) & 5.0629 & 0.7967 & 1.3322 & 0.4860 \\
v3509(4,3) & 6.2392 & 0.9050 & 1.3013 & 0.3458
\end{tabular}
\end{center}
\end{table}


\section{Topology and Cosmology}

It is presently believed that our Universe is correctly described by
a perturbed \frl~(FL) model. 
That is, homogeneous and isotropic solutions of Einstein's
equations are used, of which the spatial sections
have constant curvature. The observational fact that the Universe
is, in an exact sense, inhomogeneous and anisotropic, is modelled
by perturbing these solutions. Beside  the usual  ``big--bang"
solutions, the FL models also include   the de Sitter solution, as
well as  those
incorporating a cosmological constant,  or a non standard  equation
of state. From a
spatial point of view,  the FL models fall into 3 general classes,
according to
the sign of their spatial curvature $k=-1$, 0, or~$+1$. The
spacetime manifold is described by the \rw ~metric
\begin{equation} \label{rwmetric}
ds ^2 = c^2~dt^2 - R^2(t) ~d\sigma ^2,
\end{equation} where  $$  d\sigma ^2 = d\chi ^2 + S_k ^2 (\chi )
(d\theta ^2 +
sin^2\theta ~d\phi ^2) $$  is the metric of a 3--dimensional
homogeneous manifold,
flat [$k=0$, see eq.~(\ref{e3m})] or with curvature  [$k=\pm 1$, see
eqs~(\ref{ms3}), (\ref{h3m})]. We have defined the function  \[ \left\{
\begin{array}{ll}
  S_k(\chi) = \sinh(\chi)   & \mbox{if  $k=-1$}\\

  S_k(\chi) = \chi & \mbox{if  $k=0$}\\

  S_k(\chi) = \sin(\chi)  & \mbox{if  $k=1$}
\end{array}
\right. \]
and $R(t)$  is the scale factor, chosen equal to the spatial
curvature radius for non
 flat models, so that  $k/ R^2$  is the  spatial curvature.

The most usually considered cosmological models
are those with  simply--connected space. We will refer  generically
to them
as the Simply--Connected Models, hereafter SCM's. However, the
assumption of
simple--connectedness is  arbitrary and can be dropped  out.  The
multi--connected cosmological models
(hereafter MCM's) are those having   multi--connected
spatial
sections.  To any MCM is associated a unique SCM sharing exactly the
same  kinematics and dynamics. The universal covering (UC) of the
spatial  sections of an MCM
is the  spatial section of the corresponding SCM
(${\bf R}^3$,
${\bf H}^3$ or ${\bf S}^3$ for FL models ) at the cosmological time
defining the spatial section.  In
particular, the scale factors $R(t)$  are exactly identical. In fact,
most characteristics of the
{\frl} models are preserved when we turn to the MCM's, and this is a
precious
guide for their study.

 The values of the metric parameters [the density parameter
$\Omega_0$ and the dimensionless cosmological constant,
$\lambda_0 \equiv c^2 \Lambda/(3 H_0^2)$; in the
notation of \citet{Peeb93}
$\lambda_0 \equiv \Omega_\Lambda $]
are not known accurately enough
to
decide the sign of the curvature.
In fact, it is difficult to see how dynamical measurements can
distinguish between $\Omega_0+\lambda_0=1$
and $\Omega_0+\lambda_0=1 \pm \delta$ where $0 < \delta \ll 1.$
On the contrary, significant detection of
non--trivial topology would dictate the sign of the spatial
curvature, since the structures of the holonomy groups are completely
different for the
cases $k = 1,~0,~ -1$.
Moreover,
it is worthwhile to note that the  multi--connectedness of space would
provide precise constraints on the metric parameters independently
of the cosmic distance ladder \citep{RL99}.

Thus, beyond its own specific interest, multi--connectedness would
offer a very efficient
tool of investigation  in other aspects of observational cosmology.

In SCM's, the only scale in comoving space which is not fixed is the
curvature radius,   normally defined as the present value of the scale
factor  $R(t_0) =
R_C$ (there is  no scale at all in the flat case). Thus $R_C$  is the
natural length unit in
comoving space for the UC  of an MCM.
 The Friedmann equations imply the relation
  \begin{equation}
\Omega_0 + \lambda_0  - 1= \frac{k~c^2~ }{~ H_0^2~ R_C^2}.
\end{equation}

The value of $R_C$ remains a matter of considerable debate.
The only cosmological length to which we
have a direct observational access is  the Hubble length $$L_{Hubble}
= c
H_0^{-1}=3\,000~  h^{-1}~\mbox{Mpc}.$$  If we define  $f=\sqrt{ \mid
\Omega+\lambda-1\mid }$, we have for a non--flat universe (for a flat
universe,
the value of $R_C$ remains arbitrary):
\begin{equation}
R_C = L_{Hubble}~ f^{-1} = 3\,000 ~(f~h)^{-1}\mbox{Mpc}.
\end{equation}

In practice, the Hubble length is of the same order of
magnitude as $R_H$, the particle horizon, which
defines the observable universe as a sphere in the UC. This
is just slightly
larger than the radius to the last scattering surface, which
is the present limit to observations.

Another  natural cosmological length is associated with
the cosmological constant:
\begin{equation}
L_{\Lambda} = \sqrt{1/\Lambda} =
\sqrt{ (3 H_0^2 \lambda_0 /c^2)^{-1} } = 1730
/\sqrt{\lambda_0}~\hMpc.
\end{equation}

The concept of horizon keeps its exact validity in  the MCM's, but
must
be applied to  the {\uc} space:  an  image is potentially visible
iff  its
(comoving) distance is smaller than  $R_H$ in the \uc.

   In an MCM,   additional spatial scales are  associated with the
topology, those
of  the \fp. The geometry  suggests to   compare them with  $R_C$ but
it is often more convenient,  for  observations, to compare them to
 $R_H$, or to evaluate them in~Mpc or $\hMpc$.

Figure~\ref{f-rprmrinj} and its caption give the basic definitions.
Observable effects linked to the multi--connectedness will only occur
if these
scales are smaller than  the size of the observable   universe, \ie,
the horizon radius.

\frprmrinj

\tobsvn

Dropping the hypothesis that real space is simply--connected has various
implications.
If at least one of the topological scales is smaller than the
horizon, then this will, in principle, be observable:
multiple images
of the same object or radiation emitting region will exist.
The smaller the fundamental domain, the easier it is to observe
the multiple topological imaging.
It has recently been calculated \citet{LLU98}, for a given catalog
of observable cosmic sources (discrete or diffuse)
with a given depth in redshift,
the approximate number of topological images in locally hyperbolic and locally
elliptic spaces as a function
 of the cosmological paramaters $\Omega_0$ and  $\lambda_0$. How do
the present observational data constrain the possible
multi--connectedness
of the universe? And, more generally, what kinds of tests are
conceivable ?
The following section deals with these matters.

\section{Observational methods, candidates and constraints} \label{s-methods}

\fsizes

\subsection{``Topology'' for the observer}
The simplest observational point of view for approaching
cosmological topology is to consider the observable sphere,
in comoving coordinates, in the covering space.

For brevity, an abuse of language is often adopted where a
``topology'' is considered to mean a specific 3--manifold,
expressed with some or all of the quantitative parameters
in physical units with a definite orientation,
in some standard astronomical coordinate system.
Formally, this can be expressed as the set
\begin{equation}
\label{e-obstopology}
\{\widetilde{M}\equiv \kappa_0 \equiv \Omega_0 + \lambda_0 -1,\;
\Omega_0,\;
\Gamma,\; \delta \widetilde{M},\; \delta \Omega_0,\; \delta \Gamma \}
\end{equation}
where
$\widetilde{M}$ is the covering space, which both determines and is
determined uniquely by the curvature under the assumption of
constant curvature; $\Omega_0$ is the present-day value of
the density parameter, required (along with $\lambda_0$, which
satisfies $\lambda_0 = \kappa_0 - \Omega_0 + 1$)
in order to relate cosmological redshift $z$ and distance;
$\Gamma= \{ g_i\}$ is the holonomy group of isometric linear transformations
within the covering space which map topological images onto
one another, (e.g. combinations of
translations, rotations and reflections, if $\widetilde{M} = {\bf R}^3$);
and $\delta$ is used to indicate (in abbreviated form) that the
observational uncertainties in all the quantitative parameters
need to be determined.

An equivalent, more physical, way of thinking of this set is
in terms of the $FD$, or fundamental polyhedron, which we label $P$, i.e.
\begin{equation}
\label{e-obstop_B}
\{P \equiv \widetilde{M}/\Gamma,\;  \Omega_0,\; \Gamma,\; \delta P,\;
\delta \Omega_0,\; \delta \Gamma \}
\end{equation}
where $P$ is just the polyhedron shape, orientation and our
location within it (all represented quantitatively in physical
units and coordinates), without the generators, and $\Gamma$
lists the generators (also in physical units and coordinates).

The values of $\kappa_0$ of most observational interest are
$\kappa_0 \le 0.$
Fig.~\ref{f-sizesC} puts into
perspective the length scales corresponding to
the range in $\kappa_0$ and $\Omega_0$ [expressed
as the pair ($\Omega_0,$ $\lambda_0$)] considered consistent
with present observations.

In the cases of zero curvature, no geometric constraint exists
on how many copies of $P$ could exist within the
observable sphere. In the hyperbolic case,
values of $2\rinj/R_C \ltapprox 0.5 $ are common
for manifolds in the census available
with {\sc SnapPea} (e.g. Table~\ref{t-snappea}),
i.e. $2\rinj \ltapprox 1700${\hMpc} for the
hyperbolic covering space $\widetilde{M}$
shown. So, the constraint of the rigidity theorem in the
hyperbolic case does not prevent the possibility of multiple
topological images at redshifts considerably
smaller than $z=3$.

\subsection{Early work: $\ltapprox 100${\hMpc} }

Sections 11 and 12 of \citet{LaLu95} review many different
efforts to detect or constrain the value of $\rinj.$

Note that the different parameters expressing the size of
the Universe, $\rinj,$ $r_-$ and $r_+,$ in particular,
(Fig.~\ref{f-rprmrinj}) have
not always been carefully distinguished. For example, the fact
that we live in the disk of a spiral galaxy, in which dust
obscures most extragalactic observations, implies that many
lower limits estimated for the ``size'' of $P$ are in fact lower
limits on $r_+,$ but not on $\rinj.$

However, the constraints at
$\rinj \ltapprox 100${\hMpc} are strong. For example, close images of
the Coma cluster do not exist, except if hidden by the
galactic plane \citep{Gott80},
and large scale structure (walls, filaments,
voids) clearly exists and intuitively it is
difficult to see how the Universe could be much smaller.
So $ 2r_+ \gtapprox R_H/100$ is a fair estimate.

\subsection{Since 1993: $\gg 100${\hMpc} }

At larger scales,
with increasing amounts of data available from better telescopes,
observational work on cosmological topology has increased rapidly in
the last few years. Some methods test for specific candidates
or families of 3--manifold candidates (e.g. \citealt{BPS98}),
but several new generic methods
have been developed which avoid having to assume a specific manifold
to start off with \citep{LLL96,Rouk96,RE97,Corn98b}.

\begin{figure}
\centerline{\epsfxsize=8cm
\zzz{\epsfbox[183 244 447 428]{"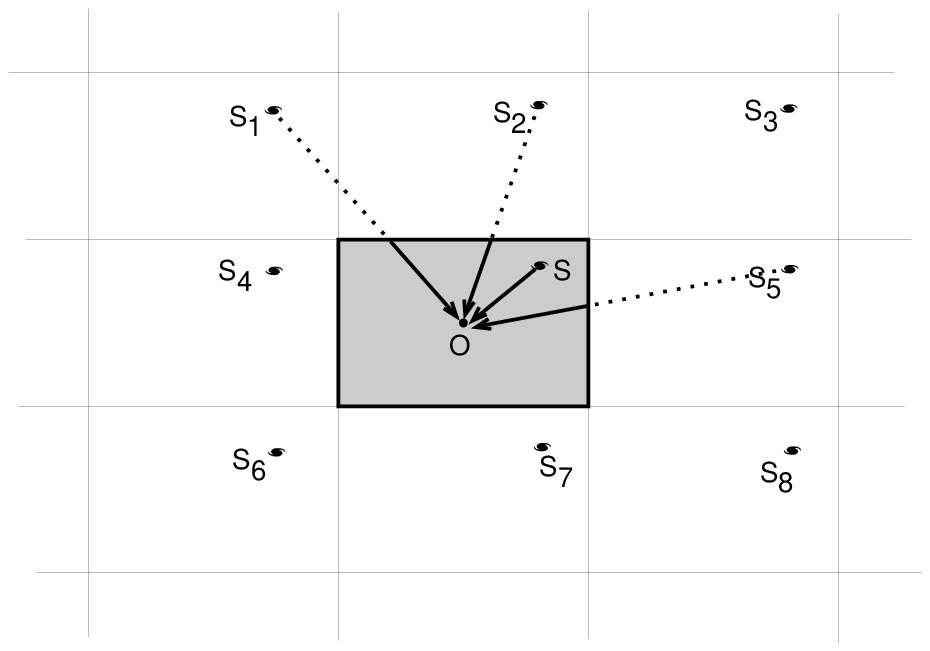"}}
{\epsfbox[183 244 447 428]{"`gunzip -c ghost.ps.gz"}}
}
\caption{Multiplication of images in the universal covering space of
a toroidal universe}
\label{ghost}
\end{figure}

The principle common to all methods is that multiple topological
images of ``objects'' should exist, if $\rinj < R_H,$ where
in the case
of the
cosmological microwave background (CMB),
regions of space whose black-body temperature can be measured
are considered as objects, though they are not
astrophysical objects in the ordinary sense. This is illustrated
for the flat torus model in Fig.~\ref{ghost}.

What varies between the methods depends on many factors.
The intrinsic
properties of the ``objects'' (or the lack of knowledge of their
properties), their intrinsic distribution throughout the observed time
cone and relative to the galactic plane, and their visibility,
make finding ``spatially well distributed, standard candles'' difficult.

Specifically, an ideal ``object'' (whether a collapsed object,
a ``configuration'' of objects, or a patch of plasma in the CMB) should:
\begin{list}{(\roman{enumi})}{\usecounter{enumi}}
\item not evolve with lookback time
\item have zero {\em three-dimensional} peculiar velocity
\item  emit isotropically
\item  be seen to large redshifts (distances)
\item  be seen over a large total volume (range in redshifts)
\item  not be obscurable by dust in the galactic plane, nor
by high latitude dust (e.g. Ophiuchus, Orion)
\end{list}

Objects which are probably the best standard candles,
i.e. clusters of galaxies detected by their hot gas in X-rays,
are mostly only seen to very small fractions of $R_H$ (though
future observations may improve this, particularly if
the curvature $\kappa_0$ is negative enough).
Objects seen to large fractions of $R_H,$ such as quasars,
are probably the worst standard candles.

This is, of course, the basic difficulty in observational
cosmology, shared by the attempts
to measure the metric parameters $\Omega_0,$
$\lambda_0$ and $H_0.$

The different methods, their advantages and disadvantages,
the claimed constraints so far, and, moreover, suggested candidates
for the 3--manifold [for ``the topology'' in the sense defined
by eq.~(\ref{e-obstopology})] are listed in
Table~\ref{t-obsvn}.
At least one of the suggested candidates
makes predictions which would be refutable
with a modest size observing programme on major telescopes.

\subsubsection{Methods: three-dimensional}

For a full review of three-dimensional observational methods, i.e.
those using astrophysical objects
distributed throughout the three-dimensional observable sphere
of the covering space, see \citet{RB98}. Only a brief description
is provided here.

\begin{figure}
\centerline{\epsfxsize=9cm
\zzz{\epsfbox[0 0 587 552]{"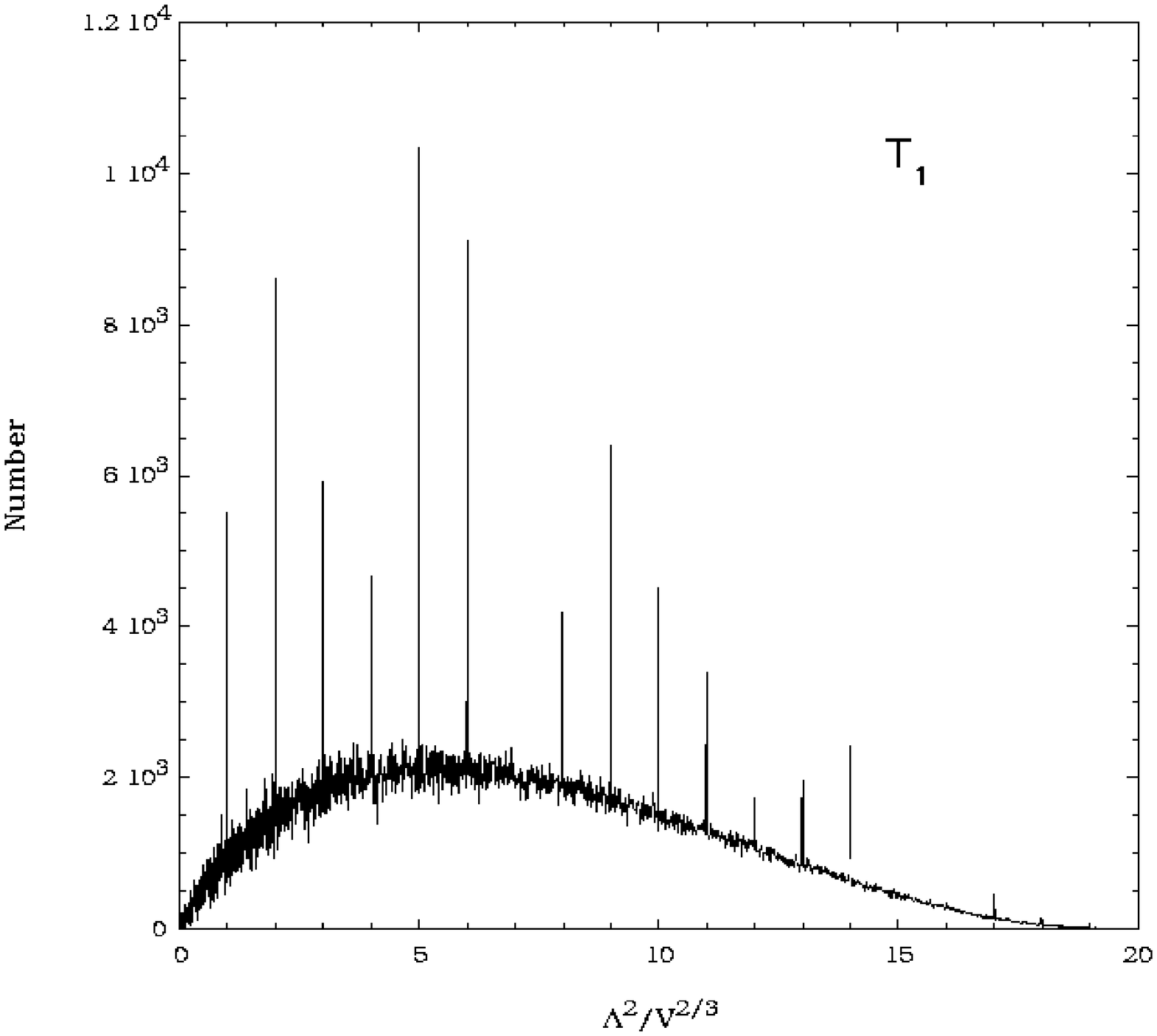"}}
{\epsfbox[0 0 587 552]{"`gunzip -c lelalu3a.eps.gz"}}
}
\caption{Histogram of pair separations for a computer generated
Einstein-de Sitter universe.  Fifty galaxies up are randomly distributed
in a cubic fundamental domain of size 2500 Mpc. Topological images are
calculated up to a redshift 4. The amplitude and relative positions
of spikes reveal the holonomies of
space and the topological type.}
\label{lelalu3a}
\end{figure}

(i) The principle behind the
method of \citet{LLL96}, ``cosmic crystallography'',
is that for good standard candles,
multiples of the vectors corresponding to the generators,
should cause very sharp peaks in a histogram of the separations
of pairs of objects (non-normalised two-point correlation function),
see Fig.~\ref{lelalu3a}.
Rich clusters of galaxies, particularly if selected in X-ray surveys,
should be fairly good standard candles at redshifts smaller than
their formation redshifts, simply because they are dominated by
hot gas in kinetic equilibrium which doesn't have time within
the age of the Universe to evolve much except to
grow by accretion and become hotter and brighter.
In addition clusters, again preferably seen
by the X-ray gas, should be relatively isotropic emitters
(compared to, say, quasars).

Lehoucq et al. \citet{LLL96} performed their
simulations in Euclidean spaces only and assumed
that the other curvature cases gave similar results. The efficiency
of the method has also been discussed by \citet{FagG98}
when the size of the physical space is comparable to the horizon
size.

Lehoucq et al. \citet{LLU98} have next analysed the
applicability of the method to elliptic and hyperbolic spaces. In the
first case there may be topological signal
in the histogram. However, small elliptical spaces are not
experimentally favored by
the present estimates of the
age of the universe. In the hyperbolic case,
no peaks in the pair 3D separations histogram were observed,
due to the fact that
the number of copies of the fundamental domain in the observable
covering space is low and that the points are not moved the same
distances by the holonomies
of space. The lack of peaks in the hyperbolic case has simultaneously
been noticed by \citet{FagG99} and \citet{Gomero99}.

Both \citet{FagG99} and \citet{ULL} have devised improvements on the
crystallographic method aimed to extract all the topological signal
the histograms, either by subtracting out the smooth (non-topological)
signal or by considering
differential histograms.

(ii) In the absence of deep, wide angle surveys for what are
the best known standard candles (galaxy clusters), other approaches
are needed.

Another use of standard candles (cf. \citealt{Gott80}) with
(most likely) monotonic evolutionary properties, i.e. clusters
seen in X-rays, is that if the highest redshift cluster in a systematic,
complete, all-sky survey is at a high redshift, then no topological
images of it can exist at lower redshifts, since they would
have to be brighter than the high redshift cluster, which in that
case would not be the brightest \citep{RE97}.

(iii) At larger scales, the objects most readily available are quasars.
These are not good standard candles, and either exist (as quasars)
only for small fractions of the age of the Universe, or recur
as bursts several times. In addition, they would be seen (except
in special cases) from different angles, and so appear as another
form of active galactic nucleus (AGN) much fainter, and not yet
catalogued.

The way around this problem \citep{Rouk96}
is to look for, in a large enough
catalogue,
the rare cases in which the evolution (and orientation) of
the quasars happens to be ideal for an Earth-based observer,
and multiple topological images are seen of several quasars
in two topological images of
a ``small region'', i.e. of a region of a few 100{\hMpc} in size.
The method is then to search (in 3-D, using redshifts for distance
estimation) for all possible isometries between
configurations of quasars in such ``small regions''.
There is not much time for the relative positions of quasars
to change much within a configuration,
so in a sense, the configurations correspond to pseudo-objects
of size 100{\hMpc} which, in a small number of cases, effectively
do not evolve.

Given contamination by chance isometries of configurations unrelated
to topology \citep{Rouk96}, theoretical work (analytical and/or
simulations) is required to find the statistically optimal
implementations of this method. For example, although weakening
the isometry criteria (e.g. matching quadruplets instead of
quintuplets) would increase the number of non-topological (noise)
isometries, it might increase the number of topological (signal)
matches by a greater factor.

(iv) A method not yet applied directly, though implicit in some
sense in the above methods, is to consider units of ``large scale
structure'' (walls, voids, etc. on scales of 50--150{\hMpc})
as objects. Although individual galaxies and quasars will be
difficult to identify at different epochs, it is expected that they
should trace out thin structures which should not move much
over the time scales involved \citep{Gott80}. One approach to analysing the
data to be taken by the X-ray satellite XMM is to use the
representations of large scale structure by the 2-dimensional
topology of contours of constant density as a characteristic
to identify potential multiple topological images of these
structures. In other words, the (2-D) topology of matter would
be used to search for the (3-D) topology of space \citep{Pierre99}.

\fcmbmethod

(v) Other methods to note include \citet{Fag85,Fag89,Fag96}
specifically for the hyperbolic case and \citet{Fag87} for looking
for images of our own galaxy as a quasar. In fact, a challenge to the
astronomers and astrophysicists who try 
understand the dynamical, inter-stellar medium (ISM) and stellar 
past history of the Milky Way (MW) and Local Group galaxies (LG) is
that they should be able to describe this history in a unique enough
fashion such that high redshift galaxies can be excluded as possible
topological images of the Milky Way just on the basis of intrinsic
properties, rather than on checking for further topological images.

Since this school is intended for doctoral (graduate) students, it
should be pointed out that this would provide a ``safe'' thesis
project with relevance for cosmic topology, in that the main goal
would be to understand the history of the MW and LG, which is already
more than enough for a thesis.  If it resulted that a ``high''
redshift galaxy (or a group) were found to have a striking resemblance
to how the MW (or LG) should have looked at that epoch, then the
student would not necessarily need to conclude that topological images
have been found --- the resemblance could be ascribed to coincidence
and checked for cosmological significance during postdoctoral work (or
by other researchers in the field).  However, the advances in the
understanding of the MW and LG history in order to make such 
a significant claim of resemblance would be sufficient for several
publishable articles.

\subsubsection{Methods: two-dimensional}

The principle behind the use of the CMB,
i.e. of a nearly two-dimensional thin shell from the observer's
point of view, is that small portions of the shell corresponding to
the CMB can be considered in some sense as objects. Depending on
the topology of the Universe, some of these ``objects'' may occupy
points
of space which have topological images on other parts of the CMB.

A schematic figure of this is shown in Fig.~\ref{f-cmbmethod}.
For the ``object'', i.e. the intersection of two sections of
the surface of last scattering (SLS),
to be seen as multiple topological images with identical
black-body temperatures, it has to have the same
average temperature integrated in different directions, and any
effects due to the
time evolution across the SLS would have to be the same in the
two directions (unless
the two effects cancelled). In other words, the CMB radiation would
have to be {\em locally isotropic}. This is expected to be
the case on large scales for the na\"{\i}ve Sachs-Wolfe effect
(gravitational redshift),
but on medium scales ($\sim 1 \deg$) the
Doppler effect, which is {\em not} isotropic, is expected to dominate.

Another possible problem with CMB multiple imaging, which has not
been explicitly analysed, is that the SLS is much thinner than the
resolution of COBE, by a factor of $\sim10^2$. That is, the black
shaded intersection in Fig.~\ref{f-cmbmethod} which is seen twice,
is averaged with $\sim 10^2$ neighbouring patches in each of two
directions perpendicular to one line of sight
to give a single observed number,
let us say in the direction of the solid arrow in the figure.
The same black patch is also observed in the direction of the dashed
arrow, but this time it is averaged with a set of $\sim10^2$ patches
in two directions perpendicular to the dashed arrow before the
single observed value is obtained.

It is not clear whether or not these two averaging processes will lead
to the same temperature fluctuation $\delta T/T$, even if the individual
patches, un-observable individually, are dominated by the same physical
process (gravitational potential). Averaging is a subtle process
in physics, and given that the averaging in this case has different
relevance depending on whether or not the hypothesis of
simple--connectedness is adopted, it is not clear that reasoning in
the former case is sufficient for the latter.

(i) The topology-independent method for using the CMB, supposing that
these two problems can be overcome,
is the ``method of circles'' \citep{Corn98b,Weeks98}.
Consider extending the tiling of the covering space by copies of
the $FD$ beyond the observable sphere and consider the observable
sphere of a second observer in one of the copies of the $FD$
outside the first sphere. The intersection of these two spheres
is a circle.

But since each copy of the $FD$ is equivalent, the
second observer can in fact perfectly well be physically
identical with the first observer. That is, in the covering
space, two observers, each
of to whom the other is behind the CMB (but not too far), are
equivalent to a single observer looking in two different
directions towards the CMB, as long as they are at the same position
within the $FD$.

Hence, the effect of multi-connectedness would be
that the values of the temperature fluctuations $\delta T/T$
around certain circles on the CMB would map onto one another,
since the circles correspond to the same set of points in
space and time.

 A related two-dimensional method to be applied to
future satellites (MAP and Planck Surveyor), is that of
searching for patterns of hot and cold spots
\citep{LevSGSB98}. This may bypass the problem of the large
number of calculations to make for the circles method.

(ii) Other two-dimensional methods require (a) assuming the topology and
specific quantitative parameters for individual 3--manifolds of
each given topology and (b) modelling the power spectrum of
perturbations. The latter is difficult to justify.

The theoretical motivation for
Gaussian amplitude distributions, uniform random phases and a
$P(k) \propto k^1$ power spectrum are unlikely to be valid at scales
approaching $\rinj$ and $r_+$. That is, either for a hyperbolic or for a
flat, $\lambda_0 \sim 0.7$ metric to be presently observable,
inflation needs some degree
of fine-tuning (which can partly be provided by the ergodicity of
geodesics in the former case, \citealt{Corn96}). Since curvature
or $\lambda_0 > 0$ must remain ``uninflated'' in the sense of
being observable at the present epoch in these cases,
it is unclear why perturbations on the scale of $R_C \sim R_H$
should necessarily be ``inflated'' in the sense of exactly satisfying
the assumptions on the power spectrum. Moreover, even for other
choices of metric, if the Universe
is observably an MCM, then scales approaching $\rinj$ and $r_+$
need not necessarily be ``inflated'' either.

The observational motivation for
Gaussian amplitude distributions, uniform random phases and a
$P(k) \propto k^1$ power spectrum are equally lacking for tests
of MCM's. The only observational justification of these properties
on large scales is COBE data, which is analysed
{\em assuming simple--connectedness}. A self-consistent assumption
on the fluctuation spectrum for comparison of an MCM with COBE
data would be to calculate this spectrum in three dimensions
based on the COBE data shifted into the $FD$ of the MCM assumed.
However, use of the resultant spectrum to simulate properties of the
COBE observations would give \ldots exactly the COBE observations
(if it has been done correctly).

So a self-consistent alternative to (b) would not enable rejection
of an MCM. In other words, this approach tests assumptions on the
the fluctuation spectrum (on the scale  $\ltapprox \rinj < r_+$)
rather than the MCM itself. (Moreover, some authors find violation of
Gaussian amplitude distributions in the COBE data
\citealt{Ferr98,Pando98}.)

Nevertheless, approach (ii) has been applied several times.
Generally, a spherical harmonic analysis (``$C_l$''
spectrum) of
simulated CMB maps (e.g. \citealt{Stev93,Star93}) is
calculated.

For a hyperbolic spatial section, a standard Fourier analysis is, in
a strict sense, invalid, whether or not a trivial topology is
assumed. This compounds the problems of assumptions regarding
the power spectrum, which cannot be defined in the usual way.
Moreover,
the classification and listing of the 3--manifolds is an open enough
problem, and knowledge of eigenmodes for the equivalent of a
Fourier analysis is even less developed.
An interesting way to avoid this problem is to use
correlation functions rather than power spectra for
assumption (b). Bond, Pogosyan \& Souradeep 
\citet{BPS98} have done this for two
of the many hyperbolic 3--manifolds known.

A general argument claimed for the methods which assume
specific 3--manifolds is that no objects, i.e. no density
fluctuations, which are larger than the Universe, i.e.
larger than $r_+$, can exist, so there should be a cutoff
at small $l$ values (large length scales) in the $C_l$
spectrum.

While this may sound common sense, it is only correct in
the $FD$. In the universal covering space, fluctuations much larger
than the Universe can indeed exist
(e.g. $\sim 30$ times larger, see Fig.~1 of \citealt{Rouk96}).
These structures cross the $FD$ many times.
While the $C_l$ spectrum may indeed lose power on large
scales due to averaging in different directions,
it is incorrect to state that fluctuations larger than
$r_+$ cannot exist. The component
eigenmodes in 3-D are of course constrained by
some integer multiple of $r_+,$ but basis functions of the 2-D
sphere cut to $\pm20\deg$ from the galactic plane only relate
to the 3-D eigenmodes via simplifying assumptions which once
again may not be valid at these length scales in MCM's.
Recently Inoue 
\citet{Inoue99} computed the low-lying eigenmodes on some compact
3-hyperbolic spaces using the direct boundary element method, and
found that the expansion coefficients behave as 
gaussian pseudo-random numbers.

A summary of the advantages and disadvantages of these
different methods is shown in the first part of Table~\ref{t-obsvn}.

Another interesting constraint may be derived when we take into
account the possibility that phase transitions  in the early universe
could develop topological defects.
Uzan \& Peter \citet{UzPet97} showed that if space
is multi--connected on scales now smaller than the horizon size,
  strings and domain walls were very unlikely to exist.
Uzan \citet{Uzan98a,Uzan98b} 
generalized these results to textures and studied the
 cosmological implications of such constraints. He concluded
that a large class of multi-connected universes with topological
defects accounting for structure formation were ruled out by the
cosmic microwave background observations.

\subsubsection{Constraints}

3-D (i,ii) The cosmic crystallography (\citealt{LLL96}, CC) and brightest
cluster (\citealt{RE97}, BC)
methods indicate lower limits on the out-diameter
of around $R_H/20$ and $R_H/10$ respectively, though the former is
valid only in
Euclidean spaces ${\bf R}^3/\Gamma$.

3-D (iii) The isometry search method applied to quasars is specifically
designed to work for poor standard candles, so is intended for
finding specific candidates for the 3--manifold, not for
constraining $\rinj$ or $r_+$.

2-D (i) No application of the identified circles principle
to COBE data to constrain $\rinj$ or $r_+$ has yet been published,
since the poor signal-to-noise and resolution make the chance of
a detection seem unlikely.

2-D (ii) However, many applications of the two-dimensional methods to
observations from the COBE satellite
\citep{Stev93,Star93,JFang94,deOliv95,LevSS98,BPS98} have been made. 
These can be interpreted either
as suggesting constraints close to the horizon scale, but only
for a (mostly Euclidean) finite number of the infinitely many topologies
possible, or
as showing that standard assumptions on the statistics of the perturbation
spectrum on scales close to $\rinj$ and $r_+$ would have to
be wrong for one of these MCM's to be correct. Note that the
confidence limits stated for several of these studies are statements
about the distribution of random realisations, not about
observational inconsistency of multiply imaged observed patches
of the CMB.

These various direct constraints and perturbation spectrum based
tests are listed in the second part of Table~\ref{t-obsvn}.

The poor resolution of COBE ($\sim 10\deg$) implies
that the data from the MAP and Planck satellites will probably
be necessary before the circles principle can be applied seriously.
This will not be easy, due to the Doppler effect, the integrated
Sachs-Wolfe effect, other foregrounds, and the problem of averaging
mentioned above.

\subsubsection{Specific candidates}

The most specific candidates suggested are those
of \citet{RE97} and \citet{BPS98}, though an implicit
candidate may also be implied by the application of the
isometry search method to quasars \citep{Rouk96}.

The former candidate was noticed by chance in a small list
of some of the brightest clusters chosen to illustrate
the brightest cluster method. Among these seven clusters,
three form a right angle to within 2\%-3\% accuracy (depending
on the values of $\Omega_0$ and $\lambda_0$) and the two
side lengths adjacent to the near-right angle are equal to
within 1\%. This is better than tolerance considered for
special quasar configurations by \citet{Fag87}.

An {\em a posteriori} estimate of the probability of this
occurring by chance would be difficult to make objectively,
but since the configuration
\begin{list}{(\roman{enumi})}{\usecounter{enumi}}
\item corresponds to that expected for the most commonly studied
candidate in CMB studies (variations on hypertori),
\item implies generators which can easily be used to derive
observationally refutable predictions,
and
\item can be used to test in detail the validity of the different
statistical methods,
\end{list}
it is useful to consider the implied
$\widetilde{M}/\Gamma = T^2 \times {\bf R}$
as a candidate 3.

See \citet{RE97} for object names, positions and
values of the generators. A more recent
summary of arguments for and
against the candidate is provided in the discussion section
of \citet{RBa99}; see also \citet{Rouk99} for a COBE analysis. 
If the candidate were correct, then two
topological images of the Coma cluster would be already known.

Bond et al. \citeauthor{BPS98} (1998, \S4.3)
have also noticed a 3--manifold (for the two
hyperbolic topologies considered) which, for a certain orientation,
is ``preferable to standard cold dark matter'' when compared
to the COBE observations. The volume of this manifold is slightly
larger
that that of the observable sphere, but $\rinj$ is less than
$R_H,$ though not enough to imply any multiple topological images
at observable sub-SLS distances.

\subsection{Other observational difficulties}

Another observational complication, which has not generally been
considered, particularly in the claims of constraints, i.e.
lower limits on $\rinj,$ is an important point by
\citet{EllisS86}: in a strict sense, it is already
well established that we do not live in an FL universe.
We live in what (locally, at least) is an almost FL universe.
Density perturbations (humans, planets, stars, galaxies, large
scale structure) exist.

Depending on the way in which space inside
the observable sphere and the $FD$ differs from
an exact FL universe, the real spatial geodesics will not
correspond exactly to multiples (and sums) of the generators
derived assuming an exact FL universe.
Gravitational lensing is one example, though the effect is probably
small.

\section{Conclusion}

The present surge of research in the mathematics, the physical
theory and the observations of cosmological topology
promise exciting developments in all three fields.
Moreover, some of the spin-offs which would follow
from a significant detection have been explicitly calculated:
the measurement of {\em transversal} galaxy velocities \citep{RBa99},
 precise estimates of the metric parameters \citep{RL99}, position of
 the topological signature in the $\Omega_{0}$--$\lambda_{0}$ plane
\citep{ULL}.

Mathematics: Continued work by mathematicians to understand and
classify
hyperbolic 3--manifolds may help theorists develop ideas
regarding the physics behind global topology, while further
developments in the software ({\sc SnapPea}) will be useful to
observational astronomers.

Theory: The development of predictions by
quantum gravity theories {\em before} a 3--manifold candidate is
confirmed to high significance would be useful from an
epistemological point of view.

Observations: The numerous observational methods offer plenty
of scope for improvements in the practical details while waiting
for new surveys such as the SDSS (Sloan Digital Sky
Survey) and new satellites such as XMM (X-ray Multiple Mission),
MAP and Planck Surveyor.

However, it is not necessary to wait: specific candidates for the
3--manifold have been and should continue to be proposed.  The
benevolence of time allocation committees for the
observational refutation (or strengthening) of the candidates
should contribute significantly to observational cosmo-topology.


\end{document}